\documentclass[11pt]{article}
\usepackage{geometry}                % See geometry.pdf to learn the layout options. There are lots.
\usepackage{pgfplots}
\usepackage{graphicx}
\usepackage{tikz}
\usepackage{marvosym}
  \usepackage{tikzsymbols}
\usetikzlibrary{patterns}
\usetikzlibrary{matrix}
\usetikzlibrary{tikzmark}
\usetikzlibrary{fit}
\usetikzlibrary{shadows.blur}
\usetikzlibrary{shapes.symbols}
\usepackage{menukeys}

\usepackage{cite}

\usepackage{mdframed}

\usetikzlibrary{positioning}
\usepackage{cancel}
\usepackage[normalem]{ulem}
\usepackage{amssymb, amsmath, amsthm}
\usepackage{epstopdf}
\newtheorem{theorem}{Theorem}[section]
\newtheorem{lemma}{Lemma}[section]
\newtheorem{definition}{Definition}[section]
\def\tina{{\mbox{\tiny \normalfont TINA}}}
\def\ctin{{\mbox{\tiny \normalfont CTIN}}}
\def\tin{{\mbox{\tiny \normalfont TIN}}}
\def\ptin{{\mbox{\tiny \normalfont P-TIN}}}
\def\sls{{\mbox{\tiny \normalfont SLS}}}
\def\bc{{\mbox{\tiny \normalfont BC}}}
\def\ic{{\mbox{\tiny \normalfont IC}}}

\newcommand{\uarrow}[1]{%
\begin{tikzpicture}[#1]%
\draw [->](0.2ex,0.5ex) -- (1ex,0.5ex)--(1ex,1ex)--(0.5ex,1ex);
\end{tikzpicture}%
}
\def\cyc{~\uarrow{scale=2}}

\title{Towards an Extremal Network Theory\\ -- Robust GDoF Gain of Transmitter Cooperation over TIN}
\author{Yao-Chia Chan, Junge Wang, Syed A. Jafar\\
{\small Center for Pervasive Communications and Computing (CPCC)}\\
{\small University of California Irvine, Irvine, CA 92697}\\
{\small \it Email: \{yaochic, jungew, syed\}@uci.edu}
}
\date{}                                           % Activate to display a given date or no date

\begin{document}

\maketitle
\allowdisplaybreaks

\begin{abstract}
Significant progress has been made recently in  Generalized Degrees of Freedom (GDoF) characterizations of wireless interference channels (IC) and broadcast channels (BC) under the assumption of finite precision channel state information at the transmitters (CSIT), especially for smaller or highly symmetric network settings. A critical barrier in extending these results to larger and asymmetric networks is the inherent combinatorial complexity of such networks. Motivated by other fields such as extremal combinatorics and extremal graph theory, we explore the possibility of an extremal network theory, i.e., a study of extremal networks within particular  regimes of interest. As our test application, we study  the GDoF benefits of transmitter cooperation in a $K$ user IC over the simple scheme of power control and treating interference as Gaussian noise (TIN) for three  regimes of interest -- a TIN regime  identified by Geng et al. where TIN was  shown to be GDoF optimal for the $K$ user interference channel, a CTIN regime  identified by Yi and Caire where the GDoF region achievable by TIN is \underline{c}onvex without time-sharing, and an SLS regime  identified by Davoodi and Jafar where a simple layered superposition (SLS) scheme is shown to be optimal in the $K$ user MISO BC, albeit only for $K\leq 3$. The SLS regime includes the CTIN regime, and the CTIN regime includes the TIN regime. As our first result, we show that under finite precision CSIT, TIN is GDoF optimal for the $K$ user IC throughout the CTIN regime. Furthermore, under finite precision CSIT, appealing to extremal network theory we obtain the following results. In the TIN regime as well as the CTIN regime, we show that the extremal GDoF gain from transmitter cooperation over TIN is bounded regardless of the number of users. In fact, the gain is exactly a factor of $3/2$ in the TIN regime, and  $2-1/K$ in the CTIN regime, for arbitrary number of users $K>1$. However, in the SLS regime, the  gain  is $\Theta(\log_2(K))$, i.e., it scales logarithmically with the number of users. %As a byproduct of our analysis we  prove a useful cyclic decomposition property of the sum GDoF achievable by TIN in the SLS regime.

\end{abstract}

{\let\thefootnote\relax\footnote{{This work is supported in part by funding from NSF grants CCF-1617504, CNS-1731384, ONR grant N00014-18-1-2057, and ARO grant W911NF-16-1-0215. This paper was presented in part at ISIT 2019.}}\addtocounter{footnote}{-1}}

\newpage
%\tableofcontents
%\newpage
%\newpage
\section{Introduction}
Finding the capacity limits of wireless networks is one of the grand challenges of network information theory. While exact capacity characterizations remain elusive, much progress has been made on this problem within the past decade through Degrees of Freedom (DoF) \cite{Zheng_Tse_Diversity} and Generalized Degrees of Freedom (GDoF) \cite{Etkin_Tse_Wang} studies. This includes both new achievable schemes, including those inspired by the idea of interference alignment (IA) \cite{Jafar_Shamai, Cadambe_Jafar_int, Motahari_Gharan_Khandani, Jafar_FnT}, and new outer bounds, such as those based on the aligned images (AI) approach \cite{Arash_Jafar}. With these  advances as stepping stones, a  worthy goal at this stage is to bring the theory closer to practice by adapting the models and metrics to increasingly incorporate practical concerns. As a step in this direction, this work is motivated by three practical concerns --- robustness, simplicity, and scalability. 

By robustness we refer specifically to the channel state information at the transmitters (CSIT). GDoF characterizations under perfect CSIT provide important theoretical benchmarks, but often lead to fragile schemes such as asymptotic \cite{Cadambe_Jafar_int} or real interference alignment \cite{Motahari_Gharan_Khandani} whose   benefits are  outweighed in practice by the potential for drastic failures due to imperfections in channel knowledge. Robustness to channel uncertainty is addressed by  GDoF characterizations that limit the CSIT to  finite\footnote{In this work by default the term GDoF will refer to GDoF under finite precision CSIT. } precision \cite{Lapidoth_Shamai_Wigger_BC}.  Optimal schemes for  such GDoF characterizations tend to be naturally robust schemes that require only a coarse knowledge of channel strength\footnote{$\alpha_{ij}$ represents the channel strength from the $j^{th}$ transmitter to the $i^{th}$ receiver, and is measured in the db scale.}    parameters $\alpha_{ij}$  at the transmitters. Aided by   advances in Aligned Images (AI) bounds \cite{Arash_Jafar}, GDoF characterizations under finite precision CSIT have been found for a variety of wireless networks in \cite{Arash_Jafar_IC, Arash_Bofeng_Jafar_BC,  Arash_Jafar_KMIMOIC, Arash_Jafar_MIMOICGDoF, Arash_Jafar_Coherence}.

The importance of simplicity is reflected in the goal  of identifying parameter regimes where  simple schemes are optimal in the GDoF sense \cite{Geng_TIN, Yi_Caire, Arash_Jafar_SLS, Sun_Jafar_ParallelTIN, Geng_Sun_Jafar_GMS, Geng_Jafar_Compound, Gherekhloo_Chaaban_Sezgin, Gherekhloo_Chaaban_Di_Sezgin, Geng_Jafar_SGDOF, Yi_Sun_TIN, Joudeh_Clerckx_MAC_TIN, Joudeh_Yi_Clerckx_BC_TIN, Yi_Sun_Jafar_Gesbert, Jafar_TIM}. The most relevant examples for our purpose are \cite{Geng_TIN}, \cite{Yi_Caire} and \cite{Arash_Jafar_SLS}. Reference \cite{Geng_TIN} identifies\footnote{While originally established under the assumption of perfect CSIT, the robustness of the TIN scheme ensures that this result carries over to finite precision CSIT. } a weak interference  regime,  
called the TIN-regime  (Definition \ref{def:tin}),  where the simple scheme of power control and treating interference as Gaussian noise (in short, TIN\footnote{Note that TIN also includes optimal power control.}) is GDoF optimal for the $K$ user interference channel (IC). A broader regime, called CTIN regime (Definition \ref{def:ctin}, the `C' signifies `convex') is identified by Yi and Caire in \cite{Yi_Caire} where, quite remarkably, the GDoF region achievable by TIN is shown to be \underline{\bf c}onvex without the need for time-sharing. It is not known whether TIN is GDoF optimal in this regime. Reference \cite{Arash_Jafar_SLS} identifies an even broader  regime, called the SLS-regime (Definition \ref{def:sls}), where a simple layered superposition (SLS) scheme is GDoF optimal for the corresponding $K$ user MISO broadcast channel (BC) under finite precision CSIT, but only for $K\leq 3$. Optimality of SLS for larger networks seems plausible, but a rapid growth in the number of parameters stands in the way of any such effort.  Comparisons between the GDoF characterizations for interference and broadcast channels in these regimes are of interest because they shed light on  the benefits of transmitter cooperation over TIN. However, based on existing results, our ability to make direct comparisons is limited to very small networks. This brings us to the third practical concern, scalability.

Wireless networks often involve a large number of users. 
Studies of large networks have to deal with an explosion in the number of parameters. One way to limit the number of parameters is to study symmetric settings. For example, consider the symmetric setting obtained by setting $\alpha_{ij}=1$ if $i=j$ and   $\alpha_{ij}=\alpha$ if $i\neq j$, for all $i,j\in[K]$.  Under finite precision CSIT, GDoF are characterized for the symmetric $K$ user interference channel in \cite{Arash_Jafar_IC}, and for the symmetric $K$ user MISO BC in \cite{Arash_Bofeng_Jafar_BC}. Based on the symmetric settings, sum-GDoF gain of the symmetric $K$ user MISO BC over the symmetric $K$ user IC is at most a factor of $3/2$ for all values of $\alpha\in[0,1]$. Furthermore, the TIN scheme can only achieve $\max(1, K(1-\alpha))$ GDoF \cite{Geng_TIN} while the $K$ user MISO BC has $\alpha+K(1-\alpha)$ GDoF \cite{Arash_Bofeng_Jafar_BC}. Therefore, transmitter cooperation can provide an improvement over TIN by a factor of at most $3/2$ in the TIN-regime and the CTIN regime (both correspond to $\alpha\leq 1/2$), and a factor of at most $2$ in the SLS-regime ($\alpha\leq 1$). Evidently the benefits of optimal transmitter cooperation over a simple scheme like TIN, are bounded for large $K$ in both regimes.

But is this also true for \emph{asymmetric} settings? To answer such questions, we need to venture beyond symmetric settings, and yet somehow avoid the curse of dimensionality. Other fields that face similar challenges, such as graph theory and set theory, find a path to progress through extremal analysis, i.e., the study of extremal graphs or extremal sets that satisfy various properties of interest. It stands to reason  that a path to progress for wireless networks may be found in extremal network theory, i.e., the study of extremal networks. This is the main idea that we wish to explore in this work. Our interest in the optimality of TIN and the benefits of transmitter cooperation provides us a  context within which we can test  the feasibility of the study of extremal networks.

We are interested specifically in the benefits of transmitter cooperation under \emph{weak interference} over the simple baseline of  TIN. The question is intriguing because on the one hand, we expect TIN to be a powerful scheme in weak interference regimes, but on the other hand full cooperation among all transmitters can also be quite powerful.  Appealing to extremal network theory, we study the  ratio, 
\begin{align}
\eta_K=\sup_{[\alpha]_{{\scriptsize K\times K}}\in\mathcal{A}}\frac{\mathcal{D}_{\Sigma,\bc}}{\mathcal{D}_{\Sigma,\tina}},\label{eq:etaK}
\end{align}
where  $\mathcal{D}_{\Sigma,\tina}$ is the supremum (maximum, if it exists) of the sum-GDoF values \emph{achievable} by power control and TIN in a $K$ user IC. This is the baseline for comparison.  $\mathcal{D}_{\Sigma,\bc}$ is the \emph{optimal} sum-GDoF of the corresponding $K$ user MISO BC obtained by full transmitter cooperation. The study of $\eta_K$ is consistent with extremal network theory because of the maximization  over $[\alpha]_{K\times K}$. Networks that maximize the ratio in \eqref{eq:etaK} are extremal networks within the class of networks specified by the regime of interest, $\mathcal{A}$. The three regimes that we consider are, the TIN-regime, $\mathcal{A}_\tin$,  the CTIN-regime, $\mathcal{A}_\ctin$, and the SLS-regime, $\mathcal{A}_\sls$.  No assumption of symmetry is made within these regimes.

As our first result (Theorem \ref{theorem:CTINTIN}), we show that under finite precision CSIT, TIN is GDoF optimal not only in the TIN regime as was already known, but also throughout the strictly larger CTIN regime for the $K$ user interference channel. Then for each of the three regimes, we characterize the extremal GDoF gain, $\eta_K$, from transmitter cooperation over TIN. For the CTIN and TIN regimes, we show that $\eta_K=\Theta(1)$, i.e., it is bounded by a constant regardless of the number of users, $K$. In fact $\eta_K=3/2$ in the TIN regime (Theorem \ref{theorem:BCICTIN}), and $\eta_K=2-1/K$ in the CTIN regime (Theorem \ref{theorem:BCICTINA}), for arbitrary number of users $K>1$. The bounded gain is consistent with and generalizes the insight obtained from the GDoF characterizations of symmetric IC and BC in \cite{Arash_Jafar_IC, Arash_Bofeng_Jafar_BC}.  For the SLS regime, we show that, $\eta_K=\Theta(\log_2(K))$, i.e., the extremal GDoF gain of transmitter cooperation over TIN grows logarithmically with the number of users (Theorem \ref{theorem:BCTINSLS}) for large networks. This is in contrast with the insights from the symmetric case where the improvement is at most by a factor of $2$. The constructive proof of this result reveals a hierarchical topology (Section \ref{sec:Nnet}) that benefits greatly from transmitter cooperation. It is also remarkable that the SLS scheme suffices to achieve the logarithmic extremal GDoF gain from transmitter cooperation over TIN.  As a byproduct of our analysis we  discover (Theorem \ref{theorem:main}) an important cyclic partition property of a TIN achievable region known as polyhedral TIN \cite{Geng_TIN} (Definition \ref{def:ptin}) that holds everywhere in the SLS-regime.

\begin{figure}[t]
\begin{center}
\begin{tikzpicture}[declare function={
    func1(\a)= (\a <= 2) * (abs(0.7*sin(x)+rand*1))   
    +              and(\a >2, \a<=5) * (1+abs(sin(x+rand)+rand*2))
 +             and(\a>5,\a<=7)*(4+2*abs(sin(x+rand))+2.5*abs(rand))
+             and(\a>7,\a<=10)*(10.8+rand*4))
+             and(\a>10, \a<=15)*(1.5+abs(sin(x+rand)+rand*3))
+       (\a>15)*(7.3+rand*2))
   ;
   func2(\a)= (\a <= 2) * (1.2)   
    +              and(\a >2, \a<=5) * (2.9)
 +             and(\a>5,\a<=7)*(6.9)
+             and(\a>7,\a<=10)*(15)
+             and(\a>10,\a<=15)*(5)
+             (\a>15)*(10)
   ;
  }]
\pgfmathsetseed{3}

        \begin{axis}[
            width=10cm, height=4cm,
            enlarge x limits=false,
            xtick=\empty,
            axis lines*=middle,
            hide y axis,
            samples=200, domain=0:20,
        ]

        \addplot [thick, black]{func1(x)};
       \addplot [red, ultra thick]{func2(x)};
        \end{axis}

    \end{tikzpicture}
    \end{center}
\caption{A conceptual depiction of  a function over a rich parameter space and its simplified representation through extremal values over various regimes of interest.}\label{fig:concept}
\end{figure}
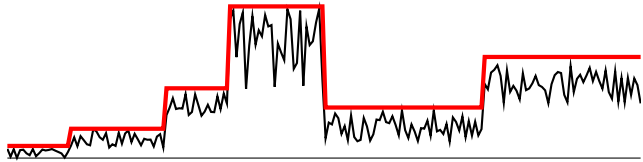

To understand the significance of these results, and of extremal network theory in general, it is important to be clear about what extremal results represent. As a visual aid, consider Figure \ref{fig:concept} where an arbitrary function is shown in black, whose rich variations make it difficult to characterize it exactly for all parameter values, and contrast it with the simpler description shown in red which bounds the \emph{range} of the function in different regimes of interest by its corresponding extremal values. 
  \noindent  The simplicity of  the extremal characterization makes it a compelling alternative  to the complexity of the complete characterization. This is also the case with GDoF characterizations for large networks, where a central challenge is the overwhelming richness of the parameter space. We similarly propose extremal network analysis as a solution to this challenge. However, in using extremal results,  it is important to remember that extremal values represent the  \emph{potential}   within each regime, and not necessarily the \emph{typical} or \emph{average} behavior. Here, let us consider two possibilities. Suppose extremal analysis shows that the potential is small, e.g., in the TIN and CTIN regimes we find that cooperation can provide at most a constant factor gain in GDoF for arbitrarily large networks, i.e., the multiplicative gain from cooperation does not scale with $K$. In fact, the constant is quite small, $1.5$ for TIN and at most $2$ for CTIN. At this point, one might reasonably conclude that the gain is too small to be be worthwhile for further studying this class of channels. Thus, small extremal values bring a measure of closure to the corresponding parameter regimes. On the other hand, surprisingly large extremal values identify regimes that merit further study. By the elephant-matchbox doctrine (elephants cannot hide in matchboxes) these are the regimes where important ideas may be discovered. It is also important to identify the extremal networks that may be studied carefully to isolate these ideas.  Last but not the least, extremal features are interesting by definition, in the same way that the speed of light, the blue whale, and Mount Everest are interesting. So whether it is intellectual curiosity, or the potential for the discovery of big ideas, or the need for a coarse understanding of overwhelmingly rich parameter spaces, the take home message of this work is that the study of extremal networks presents a promising way forward.

The paper is organized as follows. The system model appears in Section \ref{sec:model}, which also recalls a key result of \cite{Arash_Jafar_IC}, and explains the significance of  finite precision CSIT, GDoF and GDoF comparisons. Definitions of various interference regimes, cycles, partitions, and achievable regions appear in Section \ref{sec:def}. Section \ref{sec:CTINTIN} presents our first result, the optimality of TIN in the CTIN regime. Extremal GDoF gains of transmitter cooperation over TIN are then characterized for the TIN regime in Section \ref{sec:TIN}, the CTIN regime in Section \ref{sec:CTIN}, and the SLS regime in Section \ref{sec:SLS}. A conclusion appears in Section \ref{sec:conc}. Appendix \ref{app:cycpart} presents a key technical result about the optimality of cyclic partitions in the SLS regime, and Appendix \ref{app:useful} contains basic lemmas needed for the proofs of the main results. 

\emph{Notations:} For   integers $X$ and $Y$, define $[X:Y] = \{X,X+1, \cdots, Y\}$. Also define $[X] = [1:X]$. The notation $[\alpha]_{K\times K}$ represents a $K\times K$ matrix whose $(i,j)^{th}$ element is $\alpha_{i,j}$ (or $\alpha_{ij}$ if there is no potential for ambiguity).
The cardinality of a set $S$ is denoted by $|S|$.
For  functions $f(K)$ and $g(K)$, denote $f(K) = \Theta(g(K))$ if $\lim\limits_{K \rightarrow \infty} f(K) / g(K) = c$ for some finite constant $c > 0$. The notation $(x)^+$ represents $\max(x,0)$.

\section{System Model}\label{sec:model}
For GDoF studies, the $K$ user interference channel is modeled as \cite{Arash_Jafar, Arash_Jafar_IC}
\begin{align} \label{eq:model}
Y_k(t) =& \sum_{i=1}^{K} \bar{P}^{\alpha_{ki}} G_{ki}(t) X_i(t) + Z_k(t),  && \forall k \in [K].
\end{align}
During the $t^{th}$ channel use, $X_i(t), Y_k(t), Z_k(t)\in\mathbb{C}$ are, respectively, the symbol transmitted by Transmitter $i$ subject to a normalized unit transmit power constraint, the symbol received by User $k$, and the zero mean unit variance additive white Gaussian noise (AWGN) at User $k$. $\bar{P} \triangleq \sqrt{P}$, is a nominal parameter that approaches infinity to define the GDoF limit (see Section \ref{sec:GDoFsig}). The exponent $\alpha_{ki} \geq 0$ is referred to as the channel strength of the link between Transmitter $i$ and Receiver $k$, and is known to all transmitters and receivers. The channel coefficients $G_{ki}(t)$ are known perfectly to the receivers but only available to finite precision at the transmitters. The finite precision CSIT assumption implies that from the transmitter's perspective, the joint and conditional probability density functions of the channel coefficients exist and the peak values of these distributions are bounded, i.e., they do not grow with $P$ (see \cite{Arash_Jafar} for further description of the bounded density assumption). Note that the transmitters know the distributions but not the actual realizations of $G_{ki}(t)$, therefore the transmitted symbols $X_i(t)$ are independent of the realizations of $G_{ki}(t)$. In the $K$ user IC, there are $K$ independent messages, one for each user, and each message is independently encoded by its corresponding transmitter. The definitions of achievable rate tuples and capacity region, $\mathcal{C}_\ic(P)$ are standard, see e.g., \cite{Arash_Jafar}. The GDoF region of the $K$ user interference channel is defined as
\begin{align}
\mathcal{D}_\ic&=\left\{(d_k)_{k\in[K]}\left|
\begin{array}{l} 
	d_k=\lim_{P\rightarrow\infty}\frac{R_k(P)}{\log(P)},\\ 
	(R_k(P))_{k\in[K]}\in\mathcal{C}_\ic(P)
\end{array} \right. \right\}. \label{def:GDoF}
\end{align}
The maximum sum-GDoF value is denoted $\mathcal{D}_{\Sigma,\ic}$.

Allowing full cooperation among the transmitters changes the problem into a $K$ user MISO BC, where the $K$ messages are jointly encoded by all $K$ transmitters. The GDoF region for the MISO BC is denoted $\mathcal{D}_\bc$ and the maximum sum-GDoF value is denoted $\mathcal{D}_{\Sigma,\bc}$.

\subsection{Deterministic Model}\label{sec:det}
As shown in \cite{Arash_Jafar} the GDoF of the channel model in \eqref{eq:model} are bounded above by the GDoF of the corresponding deterministic model with  inputs $\bar{X}_k(t)$ and  outputs $\bar{Y}_k(t)$, defined as
\begin{align}
\bar{Y}_k(t)&=\sum_{i=1}^K\left\lfloor \bar{P}^{\alpha_{ki}-\alpha_{\max,i}}G_{ki}(t)\bar{X}_i(t)\right\rfloor,
\end{align}
where $\bar{X}_i(t) = \bar{X}_i^R(t) + j \bar{X}_i^I(t)$ with $\bar{X}_i^R(t), \bar{X}_i^I(t) \in\{0,1,2,\cdots, \lceil\bar{P}^{\alpha_{\max,i}}\rceil\}$, and $\alpha_{\max,i}=\max_{j\in[K]}\alpha_{ji}$. For all the parameter regimes considered in this work, $\alpha_{\max,i}=\alpha_{ii}$. The assumptions regarding channel coefficients $G_{ki}(t)$, channel knowledge at transmitters and receivers, and definitions of messages, codebooks, achievable rates, and GDoF are the same as before. Let us also recall a very useful bound for our current purpose, a special case of Lemma 1 in \cite{Arash_Jafar_IC}.
\begin{lemma}[Lemma 1 in \cite{Arash_Jafar_IC}]\label{lemma:AIS}
\begin{align}
&{H\left(\left.\left(\sum_{i=1}^K\lfloor \bar{P}^{\lambda_i-\alpha_{\max,i}} G_{ki}(t)\bar{X}_i(t)\rfloor\right)^{[1:T]}~\right| \mathcal{G}, W_S\right)-H\left(\left.\left(\sum_{i=1}^K\lfloor \bar{P}^{\nu_i-\alpha_{\max,i}} G_{k'i}(t)\bar{X}_i(t)\rfloor\right)^{[1:T]}~\right| \mathcal{G}, W_S\right)\notag}\\
&\leq \max_{i\in[K]}(\lambda_i-\nu_i)^+T\log(P)+T~o(\log(P)),
\end{align}
where $H(Z)$ is the entropy of $Z$, the notation $(A(t))^{[1:T]}$ stands for $(A(1), A(2), \cdots, A(T))$,  $\mathcal{G}$ is a random vector containing the values of all channel coefficients $G_{ki}(t), G_{k'i}(t)$ for $k,k',i\in[K], t\in[1:T]$,  the constants $\lambda_i, \nu_i$ are arbitrary values between $0$ and $\alpha_{\max,i}$, the set $S\subset[K]$ is an arbitrary (possibly empty) subset of users, say $S=\{i_1,i_2,\cdots,i_M\}$, and $W_S=(W_{i_1},W_{i_2},\cdots,W_{i_M})$ is comprised of the corresponding users' desired messages.
\end{lemma}
The significance of Lemma \ref{lemma:AIS} may be intuitively understood as follows. Suppose there are $K$ transmitters, transmitting symbols $\bar{X}_i(t)$, $i\in[K]$,  independent of the realizations of the bounded density channel coefficients $G_{ki}(t), G_{k'i}(t)$, for all $i,k, k'\in[K], t\in[1:T]$, and the transmitted symbols $\bar{X}_i(t)$ can be heard at two receivers, $k$ and $k'$ with power levels up to $\lambda_i$ and $\nu_i$ respectively. Then the maximum difference of entropies in the GDoF sense, that can exist between the signals received at the two receivers is no more than the maximum of the difference of the corresponding values of $\lambda_i$ and $\nu_i$ (or zero if the maximum difference is negative). In other words, the greatest  difference in the GDoF sense that can be created between the entropies of received signals at two receivers can be achieved by simply transmitting from only one antenna, which is the antenna that experiences the largest difference of channel strengths between the two receivers. Remarkably, Lemma \ref{lemma:AIS} holds for both interference and broadcast settings, i.e., the symbols $\bar{X}_i$ may be independent across $i\in[K]$ as in the IC,  or dependent as in the BC. 

It will be convenient to introduce a more compact notation for Lemma \ref{lemma:AIS}. Let us define,
\begin{align}
\mathbb{H}_g([\lambda_1, \lambda_2,\cdots,\lambda_K]\mid W_S)&\triangleq H\left(\left.\left(\sum_{i=1}^K\lfloor \bar{P}^{\lambda_i-\alpha_{\max,i}} G_{ki}(t)\bar{X}_i(t)\rfloor\right)^{[1:T]}~\right| \mathcal{G},W_S\right).\label{def:compact}
\end{align}
Using this compact notation and ignoring $o(\log(P))$ terms that are inconsequential for GDoF, the statement of Lemma \ref{lemma:AIS} becomes
\begin{align}
\mathbb{H}_g([\lambda_1, \lambda_2,\cdots,\lambda_K]\mid W_S)-\mathbb{H}_g([\nu_1, \nu_2, \cdots, \nu_K]\mid W_S)&\leq \max(\lambda_1-\nu_1,\lambda_2-\nu_2,\cdots,\lambda_K-\nu_K)^+T\log(P).
\end{align}
Note that the bounded density channel coefficients that appear in the two entropy terms in Lemma \ref{lemma:AIS}, $G_{ki}$ and $G_{k'i}$ may be different, however the $W_S$ that appears in the conditioning in both entropy terms must be the same. When Lemma \ref{lemma:AIS} is applied in the context of interference channels, the conditioning on a subset of messages allows the corresponding codeword symbols $\bar{X}_i, i\in S$ to be eliminated from the received signal, essentially by setting the corresponding $\lambda_i, \nu_i$ values to $0$, after which the conditioning on $W_S$ can be dropped because the remaining $\bar{X}_i$ are independent of $W_S$. Once the conditioning on $W_S$ is dropped, any two entropy terms may be compared and their difference  bounded by Lemma \ref{lemma:AIS}. For example, in the interference channel context, 
\begin{align}
&\mathbb{H}_g([\lambda_1,\lambda_2, \lambda_3]\mid W_2)-\mathbb{H}_g([\nu_1,\nu_2, \nu_3]\mid W_3)\notag\\
&=\mathbb{H}_g([\lambda_1,0, \lambda_3])-\mathbb{H}_g([\nu_1,\nu_2, 0])\\
&\leq\max(\lambda_1-\nu_1,-\nu_2,\lambda_3)^+T\log(P).
\end{align}
However, when Lemma \ref{lemma:AIS} is applied in the context of broadcast channels, the conditioning on $W_S$ cannot be dropped because all $\bar{X}_i$ may depend on all messages. In that case, only entropy terms conditioned on the same set of messages may be compared through Lemma \ref{lemma:AIS}. This is the main difference in how Lemma \ref{lemma:AIS} may be applied to interference and broadcast channels.
\subsection{Significance of GDoF}\label{sec:GDoFsig}
The GDoF model is essentially a generalization of the deterministic model of \cite{Avestimehr_Diggavi_Tse}. The significance of the GDoF model may be intuitively understood as follows. The channel strength parameters  represent the arbitrary and finite values of corresponding link SNRs and INRs in  dB scale for a given network setting, i.e., $\alpha_{ii}=\log(\mbox{SNR$_{ii}$})$ and $\alpha_{ij}=\log(\mbox{INR$_{ij}$})$ (see, for example \cite{Geng_TIN} for a more detailed explanation). Note that $\alpha_{ii}$ and $\alpha_{ij}$ may also be understood to be the approximate capacities of the corresponding links in isolation. Unlike the degrees of freedom (DoF) metric which proportionately scales all the transmit \emph{powers}, the GDoF model proportionately scales all the link \emph{capacities}. The  exponential scaling of powers in the GDoF model corresponds to a linear scaling of all of the corresponding link capacities by the same factor, and this factor is $\log(P)$ (note that the isolated link  with signal strength $P^{\alpha_{ij}}$ has capacity $\approx\alpha_{ij}\log(P)$, thus the scaling factor is $\log(P)$). The linear scaling of powers in the DoF model causes the ratios of capacities of any two non-zero links to approach $1$ as $P\rightarrow\infty$. Thus, a very weak channel and a very strong channel become essentially \emph{equally} strong in the DoF limit, thereby fundamentally changing the character of the original network of interest. The GDoF model on the other hand keeps  the ratios of all capacities unchanged as $P\rightarrow\infty$, so that strong channels remain strong, and weak channels remain weak. The intuition behind GDoF is that if the capacities of all the individual links in a network are scaled by the \emph{same} factor, then the overall network capacity region should scale by approximately the same factor as well --- essentially a principle of scale invariance.\footnote{While the scaling of $P$ may be interpreted as a physical scaling of transmit powers in the DoF metric (which unfortunately changes the character of the given network), $P$ does not have the same interpretation of physical transmit power in GDoF. Instead, in the GDoF setting, $P$ is just a nominal parameter, such that each value of $P$ identifies a new network according to \eqref{eq:model}. These distinct networks are lumped together by the GDoF metric based on the intuition that comes from the principle of scale invariance, i.e.,  when normalized by $\log(P)$ all of these networks should have \emph{approximately} the same capacity region (see also the discussion in \cite{Jafar_TIM}).} If so, then  normalizing by the scaling factor $\log(P)$ should produce an approximation to the capacity region of the original finite SNR network setting. This is precisely how GDoF are measured, note the normalization by $\log(P)$ in \eqref{def:GDoF}. Indeed, the validity of this intuition is borne out by numerous bounded-gap capacity approximations that have been enabled by GDoF characterizations (e.g., \cite{Suh_Tse_FB, Gou_Jafar_O1, Rini_Tuninetti_Devroye, Karmakar_Varanasi_gap, Avestimehr_Sezgin_Tse}), starting with the original result -- the capacity characterization of the $2$ user interference channel within a $1$ bit gap in \cite{Etkin_Tse_Wang}.

\subsection{Significance of Finite Precision CSIT}
Asymptotic analysis under perfect CSIT often leads to fragile schemes that are difficult to translate into practice, for example the DoF of the $K$ user interference channel have been shown in \cite{Etkin_Ordentlich, Motahari_Gharan_Khandani} to depend on whether the channels take rational or irrational values -- a distinction of no practical significance. Zero forcing schemes that rely on precise channel phase knowledge to cancel signals can fail catastrophically due to relatively small phase perturbations. Robust schemes are much more valuable in practice. Restricting the CSIT to finite precision naturally shifts the focus to robust schemes that rely primarily on a coarse knowledge of channel strengths at the transmitters. While the finite precision CSIT model \cite{Lapidoth_Shamai_Wigger_BC, Arash_Jafar} allows arbitrary fading distributions subject to bounded densities, it is instructive to consider in particular the model $G_{ki}(t)=g_{ki}^R(t)+jg_{ki}^I(t)$ where $g_{ki}^R(t),g_{ki}^I(t)$ are independent and uniformly distributed over $(1-\epsilon,1+\epsilon)$ for some arbitrarily small but positive $\epsilon$.
Interpreted this way, $G_{ki}(t)$ are seen as arbitrarily small \emph{perturbations} in the channel state that serve primarily to limit CSIT in the channel model to $\epsilon$-precision, while the coarse knowledge of channel strengths remains available to the transmitters in the form of the parameters $\alpha_{ij}$. From a GDoF perspective, these perturbations filter out fragile schemes that rely on highly precise CSIT. Indeed, the GDoF benefits of most sophisticated interference alignment  and zero forcing schemes  disappear under finite precision CSIT \cite{Arash_Jafar}. However, the benefits of robust schemes that rely only on the knowledge of channel strengths, such as  rate-splitting \cite{Rate_Splitting}, elevated multiplexing \cite{Elevated_Multiplexing}, layered superposition coding \cite{Cover72, Avestimehr_Diggavi_Tse}, and treating interference as noise \cite{Sreekanth_Veeravalli, Shang_Kramer_Chen, Motahari_Khandani, Geng_TIN} remain accessible. Thus, GDoF characterizations under finite precision CSIT provide approximately optimal solutions for power control, rate-splitting, layered superposition based schemes that are quite robust in practice. The approximately optimal solutions   serve as good initialization points for  finer numerical optimizations needed at finite SNR, and inspire approximately optimal resource allocation schemes  such as ITLinQ \cite{Naderi_Avestimehr_ITLinQ} and ITLinQ$+$ \cite{Yi_Caire_ITLinQplus}. As such GDoF characterizations under finite precision CSIT are tremendously useful in bringing theory closer to practice.

\subsection{GDoF Comparisons}
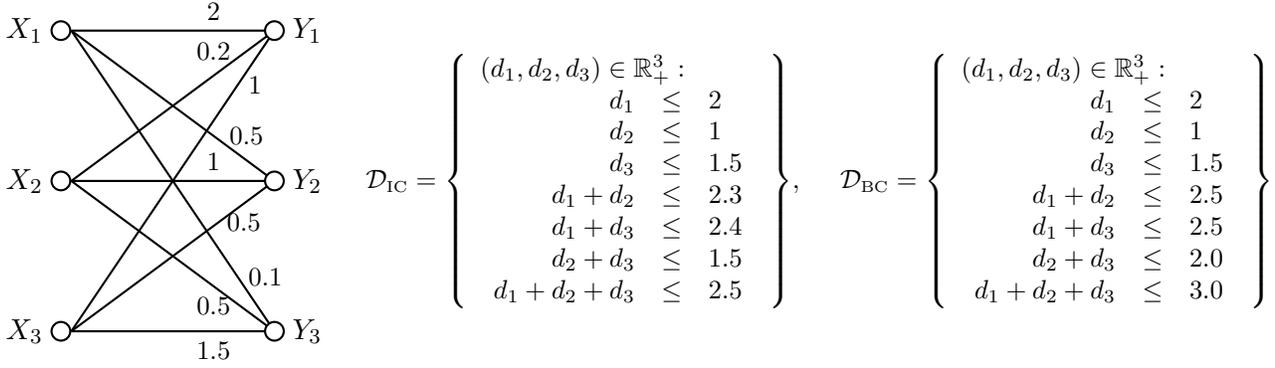
\begin{figure}[!t]
\hspace{-0.5cm}\begin{tikzpicture}
\foreach \m in {1,2,3}
{
\coordinate (M\m) at (0,1.5-2*\m);
\coordinate (N\m) at (2.7,1.5-2*\m){};
};

\draw [thick] (M1)--(N1) node[above, pos=0.7]{\small $2$};
\draw[thick] (M2)--(N1) node[above=2pt, pos=0.7]{\small $0.2$};
\draw[thick] (M3)--(N1) node[right, pos=0.82]{\small $1$};

\draw[thick] (M1)--(N2) node[right=2pt, pos=0.7]{\small $0.5$};
\draw[thick] (M2)--(N2) node[above, pos=0.7]{\small $1$};
\draw[thick] (M3)--(N2) node[below, pos=0.85]{\small $0.5$};

\draw[thick] (M1)--(N3) node[right, pos=0.82]{\small $0.1$};
\draw[thick] (M2)--(N3) node[below, pos=0.7]{\small $0.5$};
\draw[thick] (M3)--(N3) node[below, pos=0.7]{\small $1.5$};

\foreach \m in {1,2,3}
{
\node[thick, circle, draw=black, fill=white, inner sep = 2.5, left] at (M\m){};
\node[thick, circle, draw=black, fill=white, inner sep = 2.5] at (N\m){};
}
\node [left=7pt of M1] (X1) { $X_1$};
\node [left=7pt of M2] (X2) { $X_2$};
\node [left=7pt of M3] (X3) { $X_3$};
\node [right=3pt of N1] (Y1) { $Y_1$};
\node [right=3pt of N2] (Y2) { $Y_2$};
\node [right=3pt of N3] (Y3) { $Y_3$};

\node at (6.8,-2.5) {\small $\mathcal{D}_{\ic}=
\left\{\begin{array}{l}
(d_1,d_2,d_3)\in\mathbb{R}^3_+:\\
\begin{array}{rll}
d_1&\leq &2\\
d_2&\leq &1\\
d_3&\leq &1.5\\
d_1+d_2&\leq& 2.3\\
d_1+d_3&\leq& 2.4\\
d_2+d_3&\leq&1.5\\
d_1+d_2+d_3&\leq&2.5
\end{array}
\end{array}
\right\}
$,};

\node at (13.1,-2.5) {\small $\mathcal{D}_{\bc}=
\left\{\begin{array}{l}
(d_1,d_2,d_3)\in\mathbb{R}^3_+:\\
\begin{array}{rll}
d_1&\leq &2\\
d_2&\leq &1\\
d_3&\leq &1.5\\
d_1+d_2&\leq& 2.5\\
d_1+d_3&\leq& 2.5\\
d_2+d_3&\leq&2.0\\
d_1+d_2+d_3&\leq&3.0
\end{array}
\end{array}
\right\}
$};

\end{tikzpicture}
\centerline{\includegraphics[width=4in]{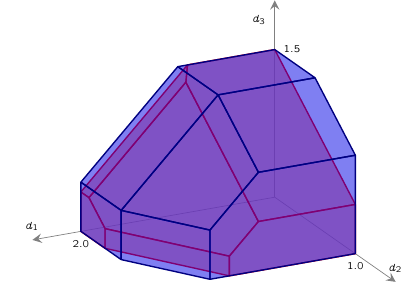}}
\caption{\it \small The GDoF region of the $3$ user interference channel in red is superimposed upon the GDoF region of the same channel with transmitter cooperation in blue. A 20$\%$ GDoF gain is seen due to transmitter cooperation for this example.}\label{fig:K3net}
\end{figure}

Comparing the GDoF of interference and broadcast channels under finite precision CSIT reveals the benefits of transmitter cooperation. As an example, consider the $3$ user interference channel with the values of $\alpha_{ij}$ parameters as shown in Fig. \ref{fig:K3net}.
The channel parameters place this setting in the TIN regime \cite{Geng_TIN}, so its GDoF region is achieved by a TIN scheme. The GDoF region is shown in red in Fig. \ref{fig:K3net}. Allowing transmitter cooperation under finite precision CSIT gives us a MISO BC. Since the TIN regime is included in the SLS regime, the GDoF of this MISO BC are characterized in \cite{Arash_Jafar_SLS}. The  GDoF region is shown in blue in Fig. \ref{fig:K3net}. Superposing the two GDoF regions we notice a significant improvement in sum-GDoF due to transmitter cooperation -- $20\%$ for this example.  We would like to perform such comparisons for larger networks, i.e., networks with more than $3$ users. However, since the results of \cite{Arash_Jafar_SLS} are limited to $3$ users, direct comparisons are not currently feasible. Instead we will explore extremal GDoF gains for large number of users. Furthermore we will limit our focus to sum-GDoF achievable by TIN and the optimal GDoF with transmitter cooperation.

\section{Definitions}\label{sec:def}
\begin{definition}[TIN Regime] \label{def:tin}\addcontentsline{toc}{subsection}{TIN Regime}
Define
\begin{align}
\mathcal{A}_\tin=\{[\alpha]_{K\times K}\in\mathbb{R}_+^{K\times K}: \alpha_{ii}\geq \alpha_{il}+\alpha_{mi} ~\forall i,l,m\in[K], i \notin \{l,m\}  \}.
\end{align}
\end{definition}
The significance of the TIN regime is that in this regime, it was shown by Geng et al. in \cite{Geng_TIN} that TIN is GDoF-optimal. 

\begin{definition}[CTIN Regime] \label{def:ctin} \addcontentsline{toc}{subsection}{CTIN Regime}
Define
\begin{align}
{\mathcal{A}}_{\mbox{\tiny \ctin}}=\{[\alpha]_{K\times K}\in\mathbb{R}_+^{K\times K}: \alpha_{ii}\geq \max(\alpha_{ij}+\alpha_{ji}, \alpha_{ik}+\alpha_{ji} -\alpha_{jk}), ~\forall i,j,k\in[K], i\notin\{j,k\}\}.
\end{align}
\end{definition}
The significance of the CTIN regime is that in this regime, it was shown by Yi and Caire in \cite{Yi_Caire} that the GDoF region achievable with TIN (also known as $\mathcal{D}_\tina$, see Definition \ref{def:tina}), is convex, without the need for time-sharing, and equal to the polyhedral TIN region over the set of all $K$ users (see Definition \ref{def:ptin}). The optimal GDoF region was heretofore unknown in the CTIN regime for the $K$ user interference channel, both under perfect CSIT and under finite precision CSIT. In this work (Theorem \ref{theorem:CTINTIN}) we settle the GDoF region in the CTIN regime under finite precision CSIT, and show that it is achieved by TIN.

\begin{figure}[h]
\centerline{\includegraphics[width=3.5in]{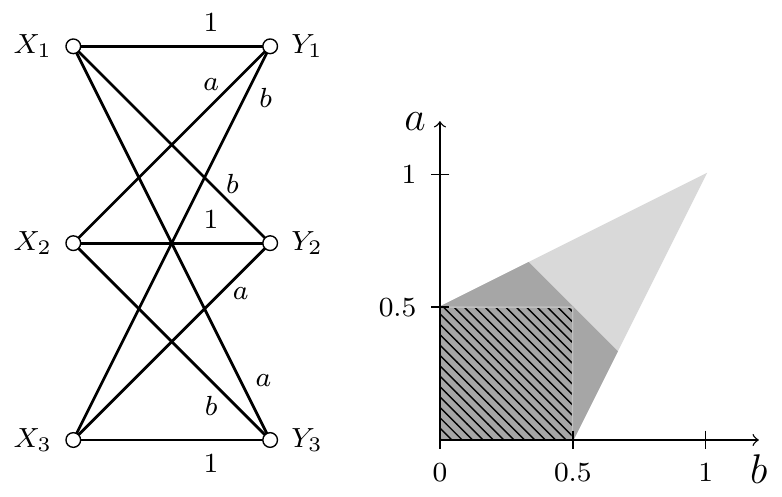}}
\caption[]{\it \small For the $3$ user symmetric setting shown here, the TIN regime is marked by the slanted line pattern, the CTIN regime includes the TIN regime and the region shaded in dark gray, and the SLS regime includes the CTIN regime and the region shaded in light gray.}\label{fig:regimes}
\end{figure}

\begin{definition}[SLS Regime] \label{def:sls} \addcontentsline{toc}{subsection}{SLS Regime}
Define the SLS regime,
\begin{align}
{\mathcal{A}}_{\mbox{\tiny SLS}}=\{[\alpha]_{K\times K}\in\mathbb{R}_+^{K\times K}: \alpha_{ii}\geq \max(\alpha_{ij},\alpha_{ki}, \alpha_{ik}+\alpha_{ji} -\alpha_{jk}), ~\forall i,j,k\in[K], i\notin\{j,k\}\}.
\end{align}
\end{definition}
The significance of the SLS regime is that in this regime, it was shown by Davoodi and Jafar in \cite{Arash_Jafar_SLS} that a simple layered superposition scheme is GDoF-optimal for the MISO BC obtained by allowing transmitter cooperation in a $K$ user interference channel. Note that the result of  \cite{Arash_Jafar_SLS} is limited to $K\leq 3$, however the regime is defined for all $K$. Also note that the SLS regime includes the CTIN regime, which includes the TIN regime. Fig \ref{fig:regimes} illustrates  the progressively larger regimes for TIN, CTIN and SLS in a $3$ user cyclically symmetric setting parameterized by channel strengths $a,b$.

\begin{definition}[Cycle $\pi$]\addcontentsline{toc}{subsection}{Cycle $(\pi)$}
A cycle $\pi$ of length $M>1$  denoted as
\begin{align}
\pi&=(i_1\rightarrow i_2\rightarrow \cdots\rightarrow  i_M\cyc~)
\end{align}
 is an ordered collection of links in the $K\times K$ interference network, that includes the desired link between Transmitter $i_m$ and Receiver $i_m$, and the interfering link between Transmitter $i_m$ and Receiver $i_{m+1}$, for all $m\in[1:M]$, where we set $i_{M+1}=i_1$, and the indices $i_1,i_2,\cdots,i_M\in[K]$ are all distinct. See Fig. \ref{fig:cycle} for an example. A cycle of length $M=1$ is  called a trivial cycle, represented simply as $\pi=(i_1\cyc~)$ for some $i_1\in[K]$, and it includes only the desired link between Transmitter $i_1$ and Receiver $i_1$. 
 \begin{figure}[h]
\centerline{\includegraphics[width=1.5in]{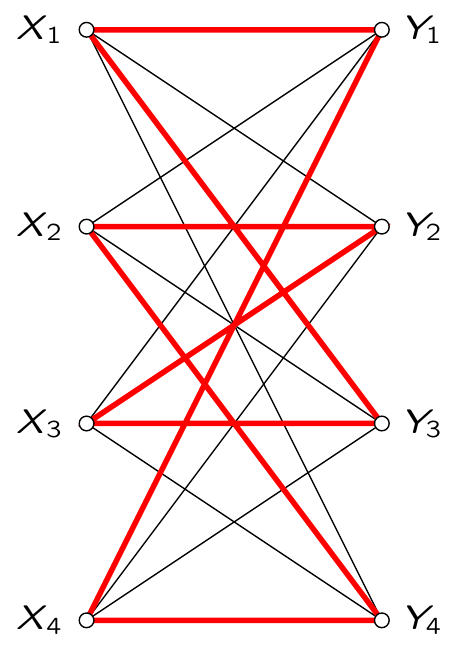}}
\caption[]{The links included in the cycle $\pi=(2\rightarrow 4\rightarrow 1\rightarrow 3\cyc~)$ are highlighted in red.}\label{fig:cycle}
\end{figure}

 Also define the following terms related to the cycle $\pi$.
\begin{enumerate}
\item Define $\pi(1)=i_1$ as the head of the cycle.  Other elements of the cycle may be similarly referenced, e.g., $\pi(2)=i_2, \pi(3)=i_3$, and so on. Thus, the cycle may be equivalently represented as $\pi=(\pi(1)\rightarrow\pi(2)\rightarrow\cdots\rightarrow\pi(M)\cyc~)$. Also note that if the cycle has length $M$, then the indices are interpreted modulo $M$, i.e., $\pi(M+i)=\pi(i)$ for all integers $i$. For example, if $\pi$ is a cycle of length $M=5$, then $\pi(6)=\pi(1), \pi(7)=\pi(2),$ etc.
\item Define $\{\pi\}=\{i_1,i_2,\cdots, i_M\}$, i.e., $\{\pi\}$ represents the set of users involved in the cycle $\pi$. 
\item Define $w(\pi)$, called the weight of the cycle $\pi$,  as the sum of strengths of all interfering links included in the cycle, i.e., $w(\pi)=\sum_{m=1}^M \alpha_{i_{m+1}i_m}$. The weight of a trivial cycle is zero because it includes no interfering links.
\item Define $\Pi$ as the set of all cycles in the $K$ user network.
\item Cycles $\pi_1,\pi_2,\cdots,\pi_n$ are said to be disjoint if the sets $\{\pi_1\},\{\pi_2\},\cdots,\{\pi_n\}$ are disjoint.
\item Cycles $\pi_1,\pi_2,\cdots,\pi_n$ are said to comprise a cyclic partition of the set $S\subset[K]$, if they are disjoint and $\bigcup_{i=1}^n\{\pi_i\}=S$.
\end{enumerate}
\end{definition}
The significance of cycles is that they lead to bounds on the sum-GDoF of the users involved in the cycle. For the interference channel, each cycle $\pi$ leads to a cycle bound $\sum_{k\in\pi}d_k\leq\Delta_\pi$ (see Definition \ref{def:Deltapi}) which is a bound on the GDoF region achievable by a restricted form of TIN, called polyhedral TIN (Definition \ref{def:ptin}). For the broadcast channel, each cycle $\pi$ leads to a bound $\sum_{k\in\pi}d_k\leq\Delta_\pi+\alpha_{\pi(i+1)\pi(i)}$ (see Lemma \ref{lemma:bctocyc} in Section \ref{sec:bctocyc}). Unlike the interference channel, the bounds for the BC are information theoretic bounds on the optimal GDoF region. These bounds are the key to all the results in this work.

\begin{definition}[Combined Cycles] \addcontentsline{toc}{subsection}{Combined Cycles}
For disjoint cycles
\begin{align}
\pi_1&=(i_1\rightarrow \cdots \rightarrow i_{M_1}\cyc~),\\
\pi_2&=(j_1\rightarrow \cdots \rightarrow j_{M_2}\cyc~),
\end{align}
the combined cycle, denoted $\pi_{1,2}=(\pi_1\rightarrow\pi_2\cyc~)$, is defined as
\begin{align}
\pi_{1,2}&=(\pi_1\rightarrow\pi_2\cyc~)=(i_1\rightarrow \cdots \rightarrow i_{M_1}\rightarrow j_1\rightarrow\cdots\rightarrow j_{M_2}\cyc~).
\end{align}
Note that $\pi_{1,2}$ is in general different from $\pi_{2,1}$.
Combinations of more than $2$ cycles are similarly defined. For example, $\pi_{1,2,3}=(\pi_1\rightarrow\pi_2\rightarrow\pi_3\cyc~)$.

\end{definition}
\begin{definition}[$\delta_{ij}$] \addcontentsline{toc}{subsection}{$\delta_{ij}$}
For $i,j\in[K]$, define
\begin{align}
\delta_{ij}&=\left\{
\begin{array}{ll}
\alpha_{ii}-\alpha_{ji},&i\neq j,\\
0, &i=j.
\end{array}
\right.
\end{align}
\end{definition}
\begin{definition}[$\Delta_\pi$] \label{def:Deltapi}\addcontentsline{toc}{subsection}{$\Delta_\pi$}
 For any cycle $\pi$ of length $M$, $\pi=(i_1\rightarrow i_2\rightarrow\cdots\rightarrow i_M\cyc~)$, define
\begin{align}
\Delta_\pi&=\left\{
\begin{array}{ll}
\delta_{i_1i_2}+\delta_{i_2i_3}+\cdots+\delta_{i_{M-1}i_M}+\delta_{i_Mi_1},&\mbox{ if $M>1$, }\\
\alpha_{i_1i_1},&\mbox{ if $M=1$}.
\end{array}
\right.
\end{align}
\end{definition}

\begin{definition}[$\mathcal{D}_\ptin(S)$] \label{def:ptin} \addcontentsline{toc}{subsection}{$\mathcal{D}_\ptin(S)$}
 For any subset of users, $S\subset[K]$, the polyhedral-TIN region \cite{Geng_TIN} is defined as
\begin{align}
\mathcal{D}_{\ptin}(S)&=
\left\{ (d_k:k\in[K]) ~ \Bigg| 
\begin{array}{ll}
0=d_k,&\forall k\in [K]\backslash S,\\
0\leq d_k,&\forall k\in S,\\
\sum_{k\in\{\pi\}}d_{k}\leq \Delta_\pi,&\forall \pi\in \Pi, \{\pi\}\subset S
\end{array} \right\}.%.
\end{align}
The bounds, $\sum_{k\in\{\pi\}}d_k\leq\Delta_\pi$, are called cycle-bounds. Note that these are not bounds on the general GDoF region, rather these are only bounds on the polyhedral TIN region for a given subset $S$. The sum-GDoF value of polyhedral-TIN over the set $S$ is defined as
\begin{align}
\mathcal{D}_{\Sigma,\ptin}(S)&=\max_{\mathcal{D}_{\ptin}(S)}\sum_{k\in S}d_k.
\end{align}
If $S=[K]$, then we will simply write $\mathcal{D}_{\Sigma,\ptin}([K])=\mathcal{D}_{\Sigma,\ptin}$.
\end{definition}
A remarkable fact about the polyhedral TIN region is that even if $S_1\subset S_2$, it is possible that the polyhedral region for $S_1$ is strictly larger  than the polyhedral region for $S_2$. See the simple example at the end of this section.

\begin{definition}[P-optimal Cyclic Partition of $S$]\addcontentsline{toc}{subsection}{P-optimal Cyclic Partition of $S$}
A cyclic partition of a subset of users $S$, $S\subset[K]$, say into the $n$ disjoint cycles $\pi_1,\pi_2,\cdots,\pi_n$, is said to be p-optimal if
\begin{align}
\mathcal{D}_{\Sigma,\ptin}(S)&=\Delta_{\pi_1}+\Delta_{\pi_2}+\cdots+\Delta_{\pi_n}.
\end{align}
\end{definition}
In general a p-optimal cyclic partition does not exist. Reference \cite{Sun_Jafar_ParallelTIN} showed that such partitions exist in the TIN regime. As one of the  key elements of this work, it is shown in Theorem \ref{theorem:main} in Appendix \ref{app:cycpart}, that such partitions must exist  in the SLS regime. Since CTIN and TIN regimes are all included in the SLS regime, these cyclic partitions exist in all three regimes.

\begin{definition}[$\mathcal{D}_\tina$] \label{def:tina}\addcontentsline{toc}{subsection}{$\mathcal{D}_\tina$}
 The TINA region  \cite{Geng_TIN, Yi_Caire} is defined as
\begin{align}
\mathcal{D}_\tina&=\bigcup_{S: S\subset[K]}\mathcal{D}_{\ptin}(S).
\end{align} 
The sum-GDoF over the TINA region are defined as
\begin{align}
\mathcal{D}_{\Sigma,\tina}&=\max_{\mathcal{D}_\tina}\sum_{k\in[K]}d_k.
\end{align}
\end{definition}
Thus the TINA region is a union of polyhedral TIN regions. In general this union does not produce a convex region.
\begin{figure}[h]
\begin{center}
\begin{tikzpicture}
\begin{scope}[shift={(1,2.5)}]
\foreach \m in {1,2,3}
{
\coordinate (M\m) at (0,1.5-2*\m);
\coordinate (N\m) at (2.7,1.5-2*\m){};
};

\draw [thick] (M1)--(N1) node[above, pos=0.5]{\small $1$};
\draw[thick] (M2)--(N1) node[above=2pt, pos=0.7]{\small $1$};

\draw[thick] (M1)--(N2) node[right=2pt, pos=0.7]{\small $1$};
\draw[thick] (M2)--(N2) node[above, pos=0.5]{\small $1$};

\foreach \m in {1,2}
{
\node[thick, circle, draw=black, fill=white, inner sep = 2.5, left] at (M\m){};
\node[thick, circle, draw=black, fill=white, inner sep = 2.5] at (N\m){};
}
\node [left=7pt of M1] (X1) { \scriptsize $X_1$};
\node [left=7pt of M2] (X2) { \scriptsize $X_2$};
\node [right=3pt of N1] (Y1) {\scriptsize $Y_1$};
\node [right=3pt of N2] (Y2) {\scriptsize $Y_2$};
\end{scope}

\begin{scope}[shift={(7,0)}]
\draw [->] (0,0)--(3,0);
\draw [->] (0,0)--(0,3);
\node at (0,0) [left, red] {\tiny $\mathcal{D}_\ptin(\{1,2\})$};
\node at (3,0) [below] {\tiny $d_1$};
\node at (2.5,0) [below] {\tiny $1$};
\node at (0,3) [left] {\tiny $d_2$};
\node at (0,2.5) [left] {\tiny $1$};
\draw [line width=1mm, blue] (0,0)--(2.5,0) node [midway, below]{\tiny $\mathcal{D}_\ptin(\{1\})$};
\draw [line width=1mm, black!30!green] (0,0)--(0,2.5) node [midway, left]{\tiny $\mathcal{D}_\ptin(\{2\})$};
\draw [fill=red](0,0) circle (1mm);
\end{scope}
\end{tikzpicture}
\caption{\small \it A $2$ user interference channel  in the SLS regime and its non-convex TINA region corresponding to the union of three polyhedral TIN regions shown in green, blue and red. }\label{fig:nonconvextina}
\end{center}
\end{figure}
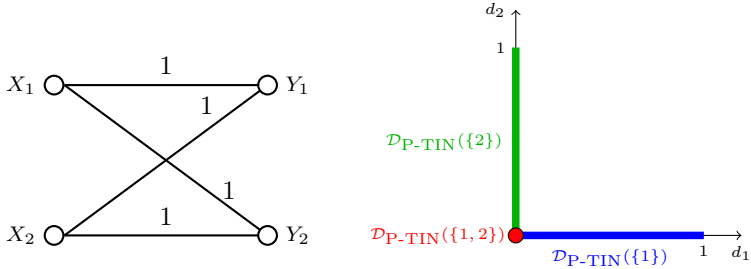
 For example, consider the $2$ user interference channel shown in Fig. \ref{fig:nonconvextina} where all $\alpha_{ij}$ values are equal to $1$. Incidentally this channel is in the SLS regime. For this channel, $\mathcal{D}_{\ptin}(\{1\})=\{(d_1,d_2): 0\leq d_1\leq 1, d_2=0\}, \mathcal{D}_{\ptin}(\{2\})=\{(d_1,d_2): d_1=0, 0\leq d_2\leq 1\}, \mathcal{D}_{\ptin}(\{1,2\})=\{(d_1,d_2): 0\leq d_1+d_2\leq 0\}=\{(d_1,d_2): d_1=0, d_2=0\}$. The union of these three regions, $\mathcal{D}_{\Sigma,\tina}=\mathcal{D}_\ptin(\{1\})\bigcup\mathcal{D}_\ptin(\{2\})\bigcup\mathcal{D}_\ptin(\{1,2\})$, is not convex. However, remarkably, the region $\mathcal{D}_\tina$ is convex for channels in the TIN regime as shown by Geng et al. in \cite{Geng_TIN}, and for channels in the CTIN regime as shown by Yi and Caire in \cite{Yi_Caire}. Next, we state our first result.

\section{TIN is GDoF optimal in the CTIN regime}\label{sec:CTINTIN}
Our first result settles the GDoF of the $K$ user interference channel in the CTIN regime. Note that all our results are under the assumption of finite\footnote{$\mathcal{D}_\ic$ remains unknown and is not always equal to $\mathcal{D}_\tina$ in the CTIN regime, if the CSIT is perfect. For example, the $3$ user IC in the CTIN regime with $\alpha_{11}=\alpha_{22}=\alpha_{33}=1, \alpha_{12}=\alpha_{23}=\alpha_{31}=1/3, \alpha_{21}=\alpha_{32}=\alpha_{31}=2/3$ achieves $3/2$ sum-GDoF by interference alignment under perfect CSIT, but TIN can achieve no more than $1$ sum-GDoF.}  precision CSIT. 
\begin{theorem}\label{theorem:CTINTIN}\addcontentsline{toc}{subsection}{Theorem \ref{theorem:CTINTIN}}
 In the CTIN regime, TIN is GDoF optimal for the $K$ user interference channel. 
\begin{align}
[\alpha]_{K\times K}\in\mathcal{A}_{\ctin} &\Rightarrow &&\mathcal{D}_{\ic}=\mathcal{D}_{\tina}.
\end{align}
\end{theorem}
\subsection{Proof of Theorem \ref{theorem:CTINTIN}}
Consider any subset of $M>1$ users, $S\subset[K]$, $|S|=M$, and let $\pi$ be a cycle of length $M$, involving these $M$ users. We will prove that the corresponding cycle bound is a valid information theoretic GDoF bound. Since we are proving an outer bound for an interference channel, without loss of generality, let us eliminate all users other than these $M$ users. This cannot hurt the  $M$ users that remain. Now, for each of the users $\pi(m), m\in[M]$, let us apply Fano's inequality within the deterministic model of the $K$ user interference channel (Section \ref{sec:det}) as follows. As usual $o(\log(P))$ terms that are inconsequential for GDoF are ignored for cleaner notation.
\begin{align}
TR_{\pi(m)}&\leq I((\bar{X}_{\pi(m)}(t))^{[1:T]}; (\bar{Y}_{\pi(m)})^{[1:T]}\mid \mathcal{G})\notag\\
&=H( (\bar{Y}_{\pi(m)})^{[1:T]}\mid \mathcal{G})-H( (\bar{Y}_{\pi(m)})^{[1:T]}\mid \mathcal{G},(\bar{X}_{\pi(m)}(t))^{[1:T]}).
\end{align}
Adding these inequalities for all $M$ users,
\begin{align}
T\sum_{m=1}^MR_{\pi(m)}&\leq \sum_{m=1}^M H( (\bar{Y}_{\pi(m)})^{[1:T]}\mid \mathcal{G})-H( (\bar{Y}_{\pi(m)})^{[1:T]}\mid \mathcal{G},(\bar{X}_{\pi(m)}(t))^{[1:T]})\\
&=\sum_{m=1}^M H( (\bar{Y}_{\pi(m)})^{[1:T]}\mid \mathcal{G})-H( (\bar{Y}_{\pi(m+1)})^{[1:T]}\mid \mathcal{G},(\bar{X}_{\pi(m+1)}(t))^{[1:T]})\\
&= \sum_{m=1}^M\Big(\mathbb{H}_g([\alpha_{\pi(m)\pi(1)}, \alpha_{\pi(m)\pi(2)},\cdots,\alpha_{\pi(m)\pi(m+1)}, \cdots, \alpha_{\pi(m)\pi(M)}])\notag\\
&\hspace{1cm}-\mathbb{H}_g([\alpha_{\pi(m+1)\pi(1)}, \alpha_{\pi(m+1)\pi(2)},\cdots,\underbrace{\cancel{\alpha_{\pi(m+1)\pi(m+1)}}}_{\mbox{\scriptsize replace with } 0}, \cdots,\alpha_{\pi(m+1)\pi(M)}])\Big)\label{eq:cancel}\\
&\leq \sum_{m=1}^M\max\Big(\max_{\ell\in[M], \ell\neq m+1}\left(\alpha_{\pi(m)\pi(\ell)}-\alpha_{\pi(m+1)\pi(\ell)}\right)^+, \alpha_{\pi(m)\pi(m+1)}\Big)T\log(P)\label{eq:AISlemma}\\
&\leq \sum_{m=1}^M\left(\alpha_{\pi(m)\pi(m)}-\alpha_{\pi(m+1)\pi(m)}\right)T\log(P)\label{eq:useCTIN}\\
&=\sum_{m=1}^M \delta_{\pi(m)\pi(m+1)}T\log(P)\\
&=\Delta_\pi T\log(P).
\end{align}
Thus, in the GDoF limit we have the bound,
\begin{align}
\sum_{m=1}^M d_{\pi(m)}&\leq \Delta_\pi.
\end{align}
Recall that  for a cycle of length $M$ the user indices are modulo $M$, i.e., $\pi(M+1)=\pi(1)$. In \eqref{eq:cancel} we used the fact that the contribution to $\bar{Y}_{\pi(m+1)}$ from $\bar{X}_{\pi(m+1)}$ can be subtracted due to the conditioning on $\bar{X}_{\pi(m+1)}$, after which the conditioning on $\bar{X}_{\pi(m+1)}$ can be dropped because in an interference channel the inputs from different transmitters are independent of each other, i.e., $\bar{X}_{\pi(m+1)}$ is independent of all remaining inputs $\bar{X}_{\pi(j)}, j\in[M], j\neq m+1$. Removing $\bar{X}_{\pi(m+1)}$ from $\bar{Y}_{\pi(m+1)}$ is equivalent to replacing the channel strength $\alpha_{\pi(m+1)\pi(m+1)}$ with zero. In \eqref{eq:AISlemma} we used the result of Lemma \ref{lemma:AIS} from Section \ref{sec:det}. In \eqref{eq:useCTIN} we used the definition of the CTIN regime, which implies that,
\begin{align}
\alpha_{\pi(m)\pi(m)}+\alpha_{\pi(m+1)\pi(\ell)}&\geq \alpha_{\pi(m+1)\pi(m)}+\alpha_{\pi(m)\pi(\ell)},\\
 \alpha_{\pi(m)\pi(m)}&\geq \alpha_{\pi(m+1)\pi(m)}+\alpha_{\pi(m)\pi(m+1)}.
\end{align}
Finally, it is trivial that for cycles of length $M=1$, the cycle bound is also an information theoretic GDoF bound. Thus, we have shown that in the CTIN regime, under finite precision CSIT, for every cycle $\pi$ in the $K$ user interference channel the cycle bound is an information theoretic GDoF bound.  The region described by these bounds is the polyhedral TIN region $\mathcal{D}_\ptin([K])$. Therefore, $\mathcal{D}_\ic\subset \mathcal{D}_\ptin([K])$. However,  $\mathcal{D}_\ptin([K])\subset\mathcal{D}_\tina$, and $\mathcal{D}_\tina\subset\mathcal{D}_\ic$. Therefore, the TIN achievable region must be the optimal GDoF region, $\mathcal{D}_\tina=\mathcal{D}_\ic$. $\hfill\square$

Next, we start presenting our results on the extremal GDoF gain from transmitter cooperation relative to TIN under the three regimes of interest -- TIN, CTIN and SLS. 

%All three regimes are weak interference regimes, where  Transmitter $i$  is the strongest possible transmitter for Receiver $i$, and vice versa. Before exploring the benefit of full transmitter cooperation, as a preliminary thought experiment suppose we only allow a re-assignment of transmitters to receivers in the $K$ user interference channel, after which only the TIN scheme is used. Because the strongest transmitters and receivers are already associated with each other, intuitively we do not expect that a reassignment would be beneficial. Indeed, this intuition is confirmed by Theorem \ref{theorem:assign} that appears in Appendix \ref{sec:assign}. In this sense, in all three regimes the natural association between Transmitter $i$ and Receiver $i$ is a \emph{stable} association.

\section{Extremal Gain from Transmitter Cooperation in  TIN Regime}\label{sec:TIN}
First, let us consider the TIN regime. Note that $K=1$ is a degenerate case because there can be no cooperation among transmitters when there is only one transmitter.
\begin{theorem}\label{theorem:BCICTIN} \addcontentsline{toc}{subsection}{Theorem \ref{theorem:BCICTIN}}
For $K\geq 2$ users,
\begin{align}
\max_{[\alpha]_{{\scriptsize K\times K}}\in\mathcal{A}_\tin}\frac{\mathcal{D}_{\Sigma,\bc}}{\mathcal{D}_{\Sigma,\ic}}&=\max_{[\alpha]_{{\scriptsize K\times K}}\in\mathcal{A}_\tin}\frac{\mathcal{D}_{\Sigma,\bc}}{\mathcal{D}_{\Sigma,\tina}}=\frac{3}{2}. \label{eq:etaKBCTIN}
\end{align}
\end{theorem}
\subsection{Proof of Theorem \ref{theorem:BCICTIN}: Upper Bound}
In the TIN regime, the GDoF of the $K$ user interference channel are achieved by TIN as shown in \cite{Geng_TIN}, so $\mathcal{D}_{\Sigma,\ic}=\mathcal{D}_{\Sigma,\tina}$.
First, let us prove the upper bound, i.e., in the TIN-regime,  $\mathcal{D}_{\Sigma,\bc}\leq 1.5\mathcal{D}_{\Sigma,\ic}$. 
Let $\pi=(i_1\rightarrow i_2\cdots \rightarrow i_M\cyc~)$ be any cycle of length $M>1$, and consider the corresponding IC cycle bound,
which is an information theoretic bound on  $\mathcal{D}_{\Sigma,\ic}(\{\pi\})$, i.e., the sum-GDoF of the IC restricted to just the users that are involved in the cycle,
\begin{align}
\mathcal{D}_{\Sigma,\ic}(\{\pi\})&\leq\delta_{i_1i_2}+\delta_{i_2i_3}+\cdots+\delta_{i_{M-1}i_M}+\delta_{i_Mi_1}= \Delta_{\pi}.
\end{align}
Note that $\Delta_{\pi}\geq \alpha_{i_1i_1}$ because $\alpha_{i_1i_1}$ GDoF are trivially achievable by simply allowing only user $i_1$ to transmit. 
For the same $M$ users, by Lemma \ref{lemma:bctocyc} in Appendix \ref{app:useful} the sum-GDoF in the BC are bounded in two ways as,
\begin{align}
\mathcal{D}_{\Sigma,\bc}(\{\pi\})&\leq \delta_{i_1i_2}+\delta_{i_2i_3}+\cdots+\delta_{i_{M-1}i_M}+\delta_{i_Mi_1}+\alpha_{i_1i_M}=\Delta_{\pi}+\alpha_{i_1i_M},\\
\mathcal{D}_{\Sigma,\bc}(\{\pi\})&\leq \delta_{i_1i_2}+\delta_{i_2i_3}+\cdots+\delta_{i_{M-1}i_M}+\delta_{i_Mi_1}+\alpha_{i_2i_1}=\Delta_{\pi}+\alpha_{i_2i_1},\\
\implies 2\mathcal{D}_{\Sigma,\bc}(\{\pi\})&\leq 2\Delta_{\pi}+\alpha_{i_2i_1}+\alpha_{i_1i_M}\leq 2\Delta_{\pi}+\alpha_{i_1i_1}\leq 3\Delta_{\pi}.\label{eq:explain}
\end{align}
In \eqref{eq:explain} we made use of the fact that in the TIN-regime, $\alpha_{i_2i_1}+\alpha_{i_1i_M}\leq\alpha_{i_1i_1}\leq \Delta_{\pi}$. 
Also for a trivial cycle, $\pi$, of length $M=1$, say comprised of only user $m$, we have $\mathcal{D}_{\Sigma,\ic}(\{\pi\})=\mathcal{D}_{\Sigma,\bc}(\{\pi\})= \alpha_{mm}=\Delta_\pi$, so here also $\mathcal{D}_{\Sigma,\bc}(\{\pi\})\leq 1.5\Delta_\pi$. Therefore for every cycle $\pi$ we have $\mathcal{D}_{\Sigma,\bc}(\{\pi\})\leq 1.5\Delta_\pi$. Now, let us consider the total GDoF of all $K$ users. Since $[\alpha]_{K\times K}\in\mathcal{A}_\tin$, from \cite{Sun_Jafar_ParallelTIN} we know that $\mathcal{D}_{\Sigma,\ic}$ is given by a cycle partition,  comprised of, say the $N$ cycles $\pi_1, \pi_2,\cdots, \pi_N$. Note that the cycles are disjoint and $\bigcup_{i=1}^n\{\pi_i\}=[K]$. 
\begin{align}
\mathcal{D}_{\Sigma,\ic}&=\sum_{n=1}^N\Delta_{\pi_n},\\
\mathcal{D}_{\Sigma,\bc}&\leq \sum_{n=1}^N \mathcal{D}_{\Sigma,\bc}(\{\pi_n\})\leq \sum_{n=1}^N1.5\Delta_{\pi_n}=1.5\mathcal{D}_{\Sigma,\ic}.
\end{align}
This completes the proof of the upper bound for Theorem \ref{theorem:BCICTIN}. $\hfill\square$
\subsection{Proof of Theorem \ref{theorem:BCICTIN}: Lower Bound}
Next, let us prove the lower bound for Theorem \ref{theorem:BCICTIN}, i.e.,  for any $K\geq 2$, there exist $[\alpha]_{K\times K}\in\mathcal{A}_\tin$, such that  $\mathcal{D}_{\Sigma,\bc}\geq 1.5\mathcal{D}_{\Sigma,\ic}$. For $K=2$ users consider the channel with $\alpha_{11}=\alpha_{22}=1, \alpha_{12}=\alpha_{21}=0.5$, for which $\mathcal{D}_{\Sigma,\ic}=1$ according to \cite{Geng_TIN} but $\mathcal{D}_{\Sigma,\bc}=1.5$ according to \cite{Arash_Jafar_cooperation}. For $K\geq 3$ it is trivial to generate such $[\alpha]_{K\times K}\in\mathcal{A}_\tin$ simply by adding trivial users $k\in[3:K]$ such that all $\alpha_{ij}$ (including the desired links $\alpha_{ii}$) associated with these additional users are zero, i.e., $\alpha_{ij} = 0$ for $i$ or $j$ is in $[3:K]$. The resulting network is still in $\mathcal{A}_\tin$.
This completes the proof of Theorem \ref{theorem:BCICTIN}.$\hfill\square$

\section{Extremal Gain from Transmitter Cooperation in  CTIN Regime}\label{sec:CTIN}

\begin{theorem}\label{theorem:BCICTINA} \addcontentsline{toc}{subsection}{Theorem \ref{theorem:BCICTINA}}
 For arbitrary number of users, $K$,
\begin{align}
 \max_{[\alpha]_{{\scriptsize K\times K}}\in\mathcal{A}_\ctin}\frac{\mathcal{D}_{\Sigma,\bc}}{\mathcal{D}_{\Sigma,\tina}}=\max_{[\alpha]_{{\scriptsize K\times K}}\in\mathcal{A}_\ctin}\frac{\mathcal{D}_{\Sigma,\bc}}{\mathcal{D}_{\Sigma,\ic}}&=2-\frac{1}{K}. \label{eq:BCICTINA}
\end{align}
\end{theorem}
Thus, the extremal GDoF gain is always less than $2$ in the CTIN regime, regardless of the number of users.
\subsection{Proof of Theorem \ref{theorem:BCICTINA}: Upper Bound}
From Theorem \ref{theorem:CTINTIN} we already know that $\mathcal{D}_{\Sigma,\tina}=\mathcal{D}_{\Sigma,\ic}$. Now let us prove the upper bound for Theorem \ref{theorem:BCICTINA}, i.e., $\mathcal{D}_{\Sigma,\bc}/\mathcal{D}_{\Sigma,\ic}\leq 2-1/K$ in the CTIN regime. For any cycle $\pi$ of length $M$, define
\begin{align}
\alpha_{\max}(\pi)&=\max_{m\in[M]}\alpha_{\pi(m)\pi(m)},\\
\alpha_{\min}(\pi)&=\left\{
\begin{array}{ll}
\min_{m\in[M]}\alpha_{\pi(m+1)\pi(m)},&M>1,\\
0,&M=1.\\
\end{array}\right.
\end{align}
In the CTIN regime, $\mathcal{D}_{\Sigma,\ptin}(\{\pi\})\leq\Delta_\pi$, and as shown by \cite{Yi_Caire}, $\mathcal{D}_{\Sigma,\ptin}(\{\pi\})\geq \mathcal{D}_{\Sigma,\ptin}(\{\pi(m)\})=\alpha_{\pi(m)\pi(m)}$ for all $m\in[M]$. Therefore,
\begin{align}
\Delta_\pi&\geq\alpha_{\max}(\pi).\label{eq:Dmax}
\end{align}
From Definition \ref{def:Deltapi},
\begin{align}
\Delta_\pi&=\sum_{m\in[M]}\alpha_{\pi(m)\pi(m)}-\alpha_{\pi(m+1)\pi(m)}\\
&\leq M\alpha_{\max}(\pi)-M\alpha_{\min}(\pi).\label{eq:Dmin}
\end{align}
From Lemma \ref{lemma:bctocyc},
\begin{align}
\mathcal{D}_{\Sigma,\bc}(\{\pi\})&\leq \Delta_\pi+\alpha_{\min}(\pi)\\
&=\Delta_\pi\left(1+\frac{\alpha_{\min}(\pi)}{\Delta_\pi}\right)\\
&\leq \Delta_\pi\left(1+\frac{\alpha_{\max}(\pi)}{\Delta_\pi}-\frac{1}{M}\right)\label{eq:useDmin}\\
&\leq \Delta_\pi\left(2-\frac{1}{M}\right).\label{eq:useDmax}
\end{align}
To obtain \eqref{eq:useDmin} we used \eqref{eq:Dmin}, and to obtain \eqref{eq:useDmax} we used \eqref{eq:Dmax}. 

Now let $\pi_1,\pi_2, \cdots, \pi_N$ be a p-optimal cyclic partition of $[K]$ into $N$ cycles of lengths $M_1, M_2, \cdots, M_N$, respectively. Then we have
\begin{align}
\mathcal{D}_{\Sigma,\bc}&\leq \sum_{n=1}^N\mathcal{D}_{\Sigma,\bc}(\{\pi_n\})\\
&\leq\sum_{n=1}^N\Delta_{\pi_n}\left(2-\frac{1}{M_n}\right)\label{eq:usegencyc}\\
&\leq\sum_{n=1}^N\Delta_{\pi_n}\left(2-\frac{1}{K}\right)\label{eq:obvious}\\
&=\mathcal{D}_{\Sigma,\ic}\left(2-\frac{1}{K}\right). \label{eq:optimalpartition}
\end{align}
\eqref{eq:usegencyc} was obtained by using \eqref{eq:useDmax}, and \eqref{eq:obvious} follows because  any cycle involves at most $K$ users, $M_n\leq K$. Finally, \eqref{eq:optimalpartition} follows because $\pi_1,\cdots,\pi_N$ represent the p-optimal cyclic partition, so $\mathcal{D}_{\Sigma,\ptin}([K])=\sum_{n=1}^N\Delta_{\pi_n}$, and because we are in the CTIN regime, according to \cite{Yi_Caire}, $\mathcal{D}_{\Sigma,\ptin}([K])=\mathcal{D}_{\Sigma,\tina}$ which is equal to $\mathcal{D}_{\Sigma,\ic}$ according to Theorem \ref{theorem:CTINTIN}. This proves the upper bound, i.e., $\mathcal{D}_{\Sigma,\bc}/\mathcal{D}_{\Sigma,\tina}=\mathcal{D}_{\Sigma,\bc}/\mathcal{D}_{\Sigma,\ic}\leq 2-1/K$ for all $[\alpha]_{K\times K}\in\mathcal{A}_\ctin$. $\hfill\square$

\subsection{Proof of Theorem \ref{theorem:BCICTINA}: Lower Bound}
Next let us prove the lower bound for Theorem \ref{theorem:BCICTINA}, i.e., there exists $[\alpha]_{K\times K}\in\mathcal{A}_\ctin$ such that $\mathcal{D}_{\Sigma,\bc}/\mathcal{D}_{\Sigma,\ic}\geq 2-1/K$. Let us define channel strength parameters as follows.  $\alpha_{ij}$ takes the value $K$ if $i=j$, and $\alpha_{ij}$ takes the  value in $[1:K-1]$ that is equivalent to $(j-i) \mod K$ when $i\neq j$. The channel strength parameter matrix can  be written explicitly as,
\begin{align}
		[\alpha]_{K \times K} = 
		\begin{bmatrix}
			K & 1 & 2 & 3 & \cdots & K-2 & K-1 \\
			K-1 & K & 1 & 2 & \cdots & K-3 & K-2 \\
			K-2 & K-1 & K & 1 & \cdots & K-4 & K-3 \\
			\vdots & \vdots & \vdots & \vdots & \ddots & \vdots & \vdots \\
			2 & 3 & 4 & 5 & \cdots & K & 1 \\
			1 & 2 & 3 & 4 & \cdots & K-1 & K  \\
		\end{bmatrix}.
\end{align}
Let us verify that $[\alpha]_{K\times K}\in\mathcal{A}_\ctin$. Due to the symmetry in this topology, it suffices to verify $\alpha_{11} \geq \alpha_{1j} + \alpha_{j1}$ for all $j \in [2:K]$, and $\alpha_{11} + \alpha_{jk} \geq \alpha_{1k} + \alpha_{j1}$ for all $j, k \in [2:K], j \neq k$.
		For $j\in[2:K]$, $\alpha_{1j} = j-1,$ and $\alpha_{j1} = K - (j-1)$, so we have $\alpha_{1j} + \alpha_{j1} = K \leq \alpha_{11}$. 
		Furthermore, since $\alpha_{jk} =  (k-j) \bmod K$, we have
		\begin{align}
			& \alpha_{11} + \alpha_{jk} - \alpha_{j1 } - \alpha_{1k} \\
			= & K +( (k-j) \bmod K) - ( K- (j-1)) - (k-1) \\
			=& ((k-j) \bmod K) - (k-j) \geq 0.
		\end{align}
Thus, the parameters are in the CTIN regime.	Next we show that $\mathcal{D}_{\Sigma,\tina}=K$. According to \cite{Yi_Caire}, in the CTIN regime we have $\mathcal{D}_{\Sigma, \tina}([K]) = \mathcal{D}_{\Sigma, \ptin}([K])$. So consider the cycle $\pi = ( 1 \rightarrow 2 \rightarrow 3 \rightarrow \cdots \rightarrow K \cyc ~)$,
		\begin{align}
			 \mathcal{D}_{\Sigma, \tina}([K])=\mathcal{D}_{\Sigma, \ptin}([K]) 
				&\leq  \Delta_\pi= \sum_{i=1}^{K} (K - (K-1)) = K.
		\end{align}
		But we also know that $\mathcal{D}_{\Sigma, \tina} \geq \alpha_{11} = K$ because it is possible to activate only user $1$ and achieve $K$ GDoF. Therefore, $\mathcal{D}_{\Sigma,\tina}=K$. Moreover, since TIN is GDoF-optimal in the CTIN regime according to Theorem \ref{theorem:CTINTIN}, we have $\mathcal{D}_{\Sigma, \ic} = K$. Finally, let us show that for the given channel strength parameters, $\mathcal{D}_{\Sigma,\bc}=2K-1$. We already know from Lemma \ref{lemma:bctocyc}, that $\mathcal{D}_{\Sigma, \bc} \leq \Delta_\pi + \alpha_{21} = 2K-1$. 
		Let us show that $2K-1$ sum-GDoF are also achievable in the MISO BC as follows. 		Let $U$ be a Gaussian codeword carrying $K-1$ GDoF, as a common message for all users.
		Let $V_i, i\in[K]$ be a codeword carrying $1$ GDoF, as a private message  for User $i$.  Let the $i^{th}$ transmit antenna send $X_i = c(\bar{P}^0 U + \bar{P}^{-(K-1)} V_i)$ where $c=\frac{1}{\sqrt{1+{P}^{-(K-1)}}}=\Theta(1)$ is a constant chosen to satisfy the input power constraint. 
		Receiver $k$ ($k\in[K]$) can decode codeword $U$ first while treating all $V_i$ as noise, because $U$ is heard with power $P^K$, and the noise floor due to all $V_i$ is no more than $P^1$. Thus, the SINR for decoding $U$ is $P^{K-1}$, which suffices because $U$ carries only $K-1$ GDoF. 
		After decoding and removing $U$ from the received signal, Receiver $k$  can decode $V_k$. This decoding is successful because $V_k$ is heard by Receiver $k$ with power $P$, while the interference from every other $V_i, i\neq k$ is received with no more than power $P^0$. Thus,  the SINR for decoding $V_k$ at Receiver $k$ is $P^1$, which suffices because $V_k$ carries only $1$ GDoF. Thus, the BC achieves a total of $(K-1)+K=2K-1$ sum-GDoF. This completes the proof of the lower bound for Theorem \ref{theorem:BCICTINA}.

$\hfill\square$

\section{Extremal Gain from Transmitter Cooperation in the SLS Regime}\label{sec:SLS}
\begin{theorem}\label{theorem:BCTINSLS} \addcontentsline{toc}{subsection}{Theorem \ref{theorem:BCTINSLS}}
\begin{align}
\sup_{[\alpha]_{{\scriptsize K\times K}}\in\mathcal{A}_\sls}\frac{\mathcal{D}_{\Sigma,\bc}}{\mathcal{D}_{\Sigma,\tina}}&=\Theta(\log(K)). \label{eq:BCTINSLS}
\end{align}
\end{theorem}
\subsection{Proof of Theorem \ref{theorem:BCTINSLS}: Upper Bound}
Let us describe an iterative procedure. Stage $\lambda$ of the procedure, $\lambda\in[0:\Lambda]$, is characterized by a subset of users, $S_\lambda\subset[K]$, a cyclic partition of $S_\lambda$ into $N_\lambda$ disjoint cycles $\pi_1^{S_\lambda},\pi_2^{S_\lambda},\cdots, \pi_{N_\lambda}^{S_\lambda}$, and a cyclic partition of $[K]$ into $N_\lambda$ disjoint cycles $\pi_1^{\lambda},\pi_2^\lambda,\cdots, \pi_{N_\lambda}^\lambda$. The  procedure stops in stage $\lambda=\Lambda$ as soon as we find $N_\lambda=1$.

Stage $0$ is the initialization stage. The procedure is initialized with the set $S_o=[K]$, the set of all users. Let $\pi_1^{S_o},\pi_2^{S_o},\cdots,\pi_{N_o}^{S_o}$ be a p-optimal cyclic partition of $S_o$ with at most one trivial cycle.  Such a partition exists and  produces the tight sum-GDoF bound for polyhedral TIN over $S_o$ so that
\begin{align}
\mathcal{D}_{\Sigma,\ptin}(S_o)&=\Delta_{\pi_1^{S_o}}+\Delta_{\pi_2^{S_o}}+\cdots+\Delta_{\pi_{N_o}^{S_o}}.
\end{align}
Choose $(\pi_1^{o},\pi_2^o,\cdots,\pi_{N_o}^o)=(\pi_1^{S_o},\pi_2^{S_o},\cdots,\pi_{N_o}^{S_o})$.
This completes the initialization stage. Note that because the p-optimal cyclic partition cannot have more than one trivial cycle, we must have $N_o\leq (K+1)/2$. If $N_o=1$, then $\Lambda=0$ and the procedure stops here. If not, then we move to the next stage.

Stage $1$ begins by defining the set of users,
\begin{align}
S_1&=\{\pi_1^o(1),\pi_2^o(1),\cdots,\pi_{N_o}^o(1)\}.
\end{align}
Let $\pi^{S_1}_1, \pi_2^{S_1},\cdots,\pi_{N_1}^{S_1}$ be a p-optimal cyclic partition of $S_1$ with at most one trivial cycle, so that
\begin{align}
\mathcal{D}_{\Sigma,\ptin}(S_1)&=\Delta_{\pi_1^{S_1}}+\Delta_{\pi_2^{S_1}}+\cdots+\Delta_{\pi_{N_1}^{S_1}}.
\end{align}
Note that these cycles only span $S_1$. For each of these cycles, $\pi_n^{S_1}$, $n\in[1:N_1]$, we will create a combined cycle, $\pi_n^{1}$ such that the $N_1$ combined cycles will be a cyclic partition of $[K]$. This is done as follows. Let us write the $n^{th}$ cycle, $\pi_n^{S_1}$, explicitly as,
\begin{align}
\pi_n^{S_1}&=(\pi_{n_1}^o(1)\rightarrow\pi_{n_2}^o(1)\rightarrow\cdots\rightarrow\pi_{n_{m_n}}^o(1)\cyc~).
\end{align}
Then the corresponding combined cycle is defined as
\begin{align}
\pi_{n}^1&=(\pi_{n_1}^o\rightarrow\pi_{n_2}^o\rightarrow\cdots\rightarrow\pi_{n_{m_n}}^o\cyc~) 
\end{align}
for $n\in[1:N_1]$. Now note that $\pi_1^1, \pi_2^1, \cdots, \pi_{N_1}^1$ span $[K]$, in fact they constitute a cyclic partition of $[K]$. This completes Stage $1$.

Note that $S_1$ has $N_o$ users, and the p-optimal cyclic partition does not have more than one trivial cycle, so we must have $N_1\leq (N_o+1)/2$. Furthermore, it follows from Lemma \ref{lemma:joincycle} that
\begin{align}
\Delta_{\pi_n^1}&\leq \Delta_{\pi_{n_1}^o}+\Delta_{\pi_{n_2}^o}+\cdots+\Delta_{\pi_{n_{m_n}}^o}+\Delta_{\pi_n^{S_1}}.
\end{align}
Summing over all $n\in[1:N_1]$ we have
\begin{align}
\Delta_{\pi_1^1}+\Delta_{\pi_2^1}+\cdots+\Delta_{\pi_{N_1}^1}&\leq \Delta_{\pi_1^o}+\Delta_{\pi_2^o}+\cdots+\Delta_{\pi_{N_o}^o}+\Delta_{\pi_1^{S_1}}+\Delta_{\pi_2^{S_1}}+\cdots+\Delta_{\pi_{N_1}^{S_1}}\\
&=\Delta_{\pi_1^o}+\Delta_{\pi_2^o}+\cdots+\Delta_{\pi_{N_o}^o}+\mathcal{D}_{\Sigma,\ptin}(S_1)\\
&\leq \Delta_{\pi_1^o}+\Delta_{\pi_2^o}+\cdots+\Delta_{\pi_{N_o}^o}+\mathcal{D}_{\Sigma,\tina}.
\end{align}
If $N_1=1$, then we set $\Lambda=1$ and the procedure stops here. If not, then we proceed to the next stage.

The procedure now simply repeats, so that at the  $(\lambda+1)^{th}$ stage we have the set of users
\begin{align}
S_{\lambda+1}&=\{\pi_1^{{\lambda}}(1),\pi_2^{{\lambda}}(1), \cdots, \pi_{N_\lambda}^{{\lambda}}(1)\}.
\end{align}
A p-optimal cyclic partition of $S_{\lambda+1}$ with at most one trivial cycle produces $N_{\lambda+1}$ disjoint cycles, $\pi^{S_{\lambda+1}}_1,\pi^{S_{\lambda+1}}_2,\cdots,\pi^{S_{\lambda+1}}_{N_{\lambda+1}}$, such that the $l^{th}$ cycle in this partition,
\begin{align}
\pi^{S_{\lambda+1}}_l&=(\pi^{{\lambda}}_{l_1}(1)\rightarrow \pi^{{\lambda}}_{l_2}(1)\rightarrow\cdots\rightarrow \pi^{{\lambda}}_{l_{m_l}}(1)\cyc~)
\end{align}
produces the $l^{th}$ combined cycle 
\begin{align}
\pi^{\lambda+1}_l&=(\pi^{{\lambda}}_{l_1}\rightarrow \pi^{{\lambda}}_{l_2}\rightarrow\cdots\rightarrow \pi^{{\lambda}}_{l_{m_l}}\cyc~)
\end{align}
for $l\in[1:N_{\lambda+1}]$. This completes Stage $\lambda+1$. Since $S_{\lambda+1}$ has $N_\lambda$ users, and the p-optimal cycle cannot have more than one trivial cycle, we must have $N_{\lambda+1}\leq (N_\lambda+1)/2$. Furthermore, it follows from Lemma \ref{lemma:joincycle} that
\begin{align}
\Delta_{\pi_1^{\lambda+1}}+\Delta_{\pi_2^{\lambda+1}}+\cdots+\Delta_{\pi_{N_{\lambda+1}}^{\lambda+1}}&\leq \Delta_{\pi_1^{\lambda}}+\Delta_{\pi_2^{\lambda}}+\cdots+\Delta_{\pi_{N_{\lambda}}^{\lambda}}+\mathcal{D}_{\Sigma,\tina}.
\end{align}
If $N_{\lambda+1}=1$, then the procedure stops and $\Lambda=\lambda+1$, otherwise the procedure continues. This completes the description of the procedure.

$\Lambda$ can be bounded by  using  $N_{\lambda+1}\leq (N_\lambda+1)/2$, $N_o\leq (K+1)/2$ and $N_{\Lambda-1}\geq 2$, as follows. $N_{\Lambda-1}\geq 2\Rightarrow N_{\Lambda-2}\geq 3\Rightarrow N_{\Lambda-3}\geq 5\Rightarrow\cdots\Rightarrow N_o\geq 2^{\Lambda-1}+1\Rightarrow K\geq 2^\Lambda+1\Rightarrow \Lambda\leq \log_2(K-1).$

Finally, we complete the proof of the upper bound as follows.
\begin{align}
\mathcal{D}_{\Sigma,\tina}&\geq \mathcal{D}_{\Sigma,\ptin}(S_o)\\
&=\Delta_{\pi_1^o}+\Delta_{\pi_2^o}+\cdots+\Delta_{\pi_{N_o}^o}\\
&\geq \Delta_{\pi_1^1}+\Delta_{\pi_2^1}+\cdots+\Delta_{\pi_{N_1}^1}-\mathcal{D}_{\Sigma,\tina}\\
&\geq \Delta_{\pi_1^2}+\Delta_{\pi_2^2}+\cdots+\Delta_{\pi_{N_2}^2}-2\mathcal{D}_{\Sigma,\tina}\\
&\vdots\notag\\
&\geq \Delta_{\pi_1^\Lambda}-\Lambda \mathcal{D}_{\Sigma,\tina}\\
&\geq \mathcal{D}_{\Sigma,\bc}-\mathcal{D}_{\Sigma,\tina}-\Lambda \mathcal{D}_{\Sigma,\tina},
\end{align}
where in the last step we used Lemma \ref{lemma:bctocyc}.
Substituting the bound for $\Lambda$ we obtain
\begin{align}
\frac{\mathcal{D}_{\Sigma,\bc}}{\mathcal{D}_{\Sigma,\tina}}&\leq 2+\log_2(K-1)\\
&=\Theta(\log_2(K)),
\end{align}
and the proof of the upper bound is complete. $\hfill\square$

\subsection{Proof of Theorem \ref{theorem:BCTINSLS}: Lower Bound}\label{sec:Nnet}
\begin{figure}[t]
	\centering
	\hspace{-1cm}\includegraphics[width=6.1in]{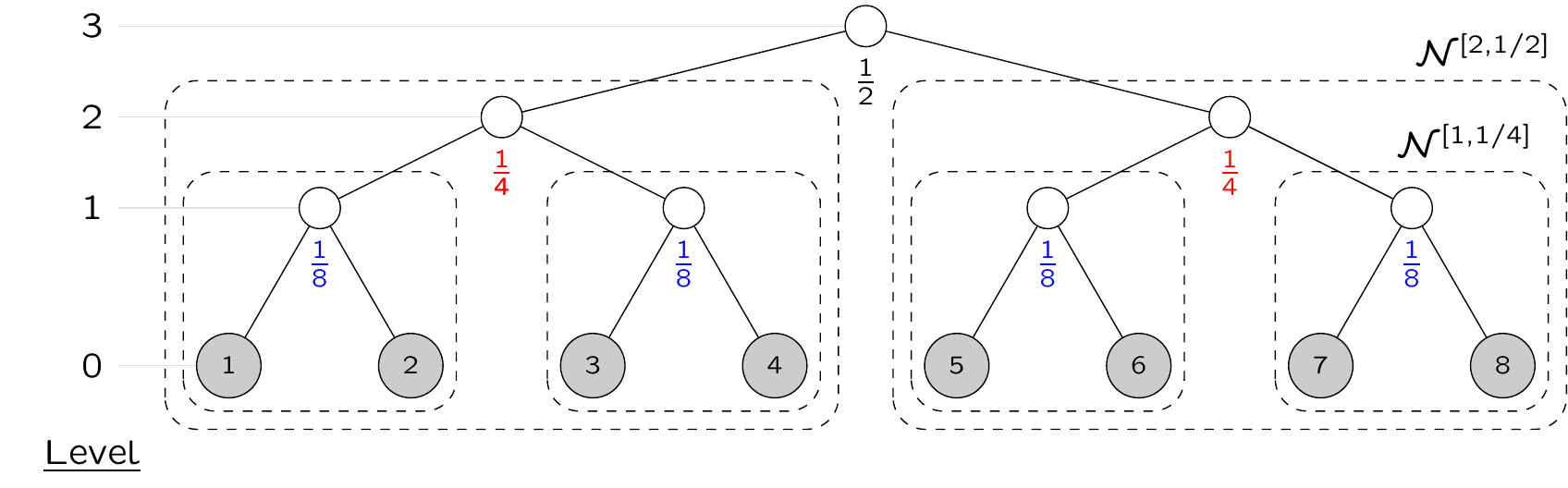}
	\caption{\small \it The binary tree representation of the network $\mathcal{N}^{[3,1]}$, and its subnetworks. The value of $\delta_{ij}^{[n,1]}=1-\alpha_{ij}^{[n,1]}$ between users $i$ and $j$ is given by the number indicated under their closest common ancestor. For example, $\delta_{78}=\delta_{87}=1/8, \delta_{14}=\delta_{41}=1/4, \delta_{37}=\delta_{73}=1/2$.}
	\label{fig:tree}
\end{figure}

For the lower bound, let us define a class of interference networks, $\mathcal{N}^{[n,\nu]}$, that is parameterized by the two numbers, $n\in\mathbb{N}, \nu\in\mathbb{R}$, $0\leq \nu\leq 1$. The number of users $K(n)=2^n$, all  desired channel strengths $\alpha_{kk}=1$, and  cross-channel strengths satisfy $\alpha_{ij}^{[n,\nu]}=\alpha_{ji}^{[n,\nu]}$ for all $i,j,k\in[K(n)]$. Since $\alpha_{ij}^{[n,\nu]}=\alpha_{ii}-\delta_{ji}^{[n,\nu]}=1-\delta_{ji}^{[n,\nu]}=1-\delta_{ij}^{[n,\nu]}$, it suffices to specify the $\delta_{ij}^{[n,\nu]}$ values instead of the $\alpha_{ij}^{[n,\nu]}$ values. To specify the $\delta_{ij}^{[n,\nu]}$ values it  will be useful to represent $\mathcal{N}^{[n,\nu]}$ as a full binary tree of depth $n$. The $2^n$ leaf nodes of this tree represent the $2^n$ users. The value of $\delta_{ij}^{[n,\nu]}=\delta_{ji}^{[n,\nu]}=\left(\frac{2^{p-1}}{2^n}\right)\nu$ if the closest common ancestor of user $i$ and user $j$ is $p$ levels above them. For example, $\delta_{ij}^{[n,\nu]}=\frac{\nu}{2^{n}}$ if user $i$ and $j$ are siblings (share a common parent), $\frac{2\nu}{2^n}$ if they share the same grandparent (but not the same parent), and the largest possible value of $\delta_{ij}^{[n,\nu]}$ in $\mathcal{N}^{[r,\nu]}$ is $\nu/2$, between users whose closest common ancestor is the root node. An interference network with these parameter values is said to be an $\mathcal{N}^{[n,\nu]}$ network. 
Fig. \ref{fig:tree} shows the binary tree for the network $\mathcal{N}^{[3,1]}$. We are primarily interested in the network for $\nu=1$. \footnote{ Even though we are interested primarily in $\nu=1$, the network $\mathcal{N}^{[n,\nu]}$ is defined for arbitrary $\nu$ because the network has a hierarchical structure and the two parameters, $n$ and $\nu$, can be used to specify the subnetworks in the hierarchy.  For example, the $\mathcal{N}^{[3,1]}$ network in Fig.  \ref{fig:tree} consists of two $\mathcal{N}^{[2, 1/2]}$ subnetworks, and each of them in turn contains two $\mathcal{N}^{[1,1/4]}$ subnetworks. These subnetworks are important for the proof of achievability (see e.g., \eqref{eq:subnet}). }

Let us first prove that  an $\mathcal{N}^{[n,\nu]}$ network is indeed in the SLS regime.  From the definition of $\delta_{ij}^{[n,\nu]}=1-\alpha_{ij}^{[n,\nu]}$, we have
\begin{align}
\alpha_{ij}^{[n,\nu]}&=1-\left(\frac{2^{p_{ij}-1}}{2^n}\right)\nu,\\
\alpha_{ki}^{[n,\nu]}&=1-\left(\frac{2^{p_{ki}-1}}{2^n}\right)\nu.
\end{align}
Since $\alpha_{ii}=1$ and $\nu\geq 0$, it is trivially verified that $\alpha_{ii}\geq\max(\alpha_{ij}(\nu),\alpha_{ki}(\nu))$ for all $i,j,k\in[K(n)]$. Now, if users $i,j$ have their closest common ancestor $p_{ij}$ levels above them, and if users $i,k$ have their closest common ancestor $p_{ki}$ levels above them, then the users $j,k$ must have a common ancestor no more than $\max(p_{ij},p_{ki})$ levels above them. Therefore,
\begin{align}
\alpha_{jk}^{[n,\nu]}&\geq 1-\left(\frac{2^{\max(p_{ij},p_{ki})-1}}{2^n}\right)\nu\\
\implies \alpha_{ii}+\alpha_{jk}^{[n,\nu]}&\geq 1+1-\left(\frac{2^{\max(p_{ij},p_{ki})-1}}{2^n}\right)\nu\\
&\geq 1+1-\left(\frac{2^{p_{ij}-1}}{2^n}+\frac{2^{p_{ki}-1}}{2^n}\right)\nu\\
&=\alpha_{ij}^{[n,\nu]}+\alpha_{ki}^{[n,\nu]}.
\end{align}
Thus the SLS condition is satisfied. 

Next we will prove that the TINA region for this network does not allow more than $2$ sum-GDoF. For this let us go through the following three steps.
\begin{enumerate}
\item
The main argument for this proof is recursive, where we repeatedly reduce a network into its subnetworks. In particular, we are interested in the \emph{left} and \emph{right} subnetworks of $\mathcal{N}^{[n,\nu]}$, as described next. Consider the root node of the binary tree representation of $\mathcal{N}^{[n,\nu]}$. It has two child nodes, say labeled as `left' and `right'. If the root node is  eliminated, then the tree splits into two binary trees, and each of those original child nodes becomes the root node of one of those trees. Let us denote these two networks as  Left$(\mathcal{N}^{[n,\nu]})$ and Right$(\mathcal{N}^{[n,\nu]})$. Let us show that each of the networks {\normalfont Left}$(\mathcal{N}^{[n,\nu]})$ and {\normalfont Right}$(\mathcal{N}^{[n,\nu]})$ is an $\mathcal{N}^{[n-1,\nu/2]}$ network, as follows.
Since the original root node is eliminated, it is obvious that  the binary tree representation of each of these subnetworks has depth $n-1$, and correspondingly each subnetwork has $2^{n-1}$ users. The channel strengths are the same as before, but since the value of $n$ has changed to $n-1$, the value of $\nu$ needs to change to $\nu/2$ to preserve the channel strengths, so in the new subnetworks we have
\begin{align}
\delta_{ij}^{[n-1,\nu/2]}&=\left(\frac{2^{p-1}}{2^{n-1}}\right)\left(\frac{\nu}{2}\right)=\left(\frac{2^{p-1}}{2^{n}}\right)\nu=\delta_{ij}^{[n,\nu]}.\label{eq:subnet}
\end{align}
where either both $i,j$ belong to the left subnetwork or both belong to the right subnetwork.

\item Next we show that $\mathcal{D}_{\Sigma,\tina}^{[n,\nu]}\leq\max\left(1,\frac{1}{2}\mathcal{D}_{\Sigma,\tina}^{[n,2\nu]}\right)$, where  $\mathcal{D}_{\Sigma,\tina}^{[n,\nu]}$ represents the optimal sum-GDoF value over the $\mathcal{D}_{\tina}^{[n,\nu]}$ region for $\mathcal{N}^{[n,\nu]}$. This is proved as follows. From Definition \ref{def:tina}, we know that $\mathcal{D}_{\Sigma,\tina}^{[n,\nu]}$ is equal to $\mathcal{D}_{\Sigma,\ptin}^{[n,\nu]}(S)$ for some subset of users, $S\subset [K(n)]$. From Theorem \ref{theorem:main} we know that $\mathcal{D}_{\Sigma,\ptin}^{[n,\nu]}(S)$ is determined by the cycle bounds corresponding to a p-optimal cyclic partition of $S$. There are two possibilities --- either the cyclic partition includes a trivial cycle, or it does not, and we will consider them one by one. 

First, suppose the p-optimal cyclic partition of $S$ does not include any trivial cycles. In that case, let $\pi=(i_1\rightarrow \cdots\rightarrow i_M\cyc~)$ be any cycle from the p-optimal cyclic partition of $S$. By assumption, the length  of $\pi$ is $M>1$. The cycle bound corresponding to $\pi$ for $\mathcal{D}_{\Sigma,\tina}^{[n,\nu]}$ is 
\begin{align}
\sum_{k\in\{i_1,\cdots,i_M\}}d_k&\leq \delta_{i_1i_2}^{[n,\nu]}+\cdots+\delta_{i_{M-1}i_M}^{[n,\nu]}+\delta_{i_{M}i_1}^{[n,\nu]}\\
&=\frac{1}{2}\left( \delta_{i_1i_2}^{[n,2\nu]}+\cdots+\delta_{i_{M-1}i_M}^{[n,2\nu]}+\delta_{i_{M}i_1}^{[n,2\nu]}\right).
\end{align}
Therefore, all the non-trivial cycle bounds $\mathcal{D}_{\Sigma,\tina}^{[n,\nu]}$ are exactly half as large as the corresponding cycle bounds in $\mathcal{D}_{\Sigma,\tina}^{[n,2\nu]}$, proving that in this case $\mathcal{D}_{\Sigma,\tina}^{[n,\nu]}=\frac{1}{2}\mathcal{D}_{\Sigma,\tina}^{[n,2\nu]}$.

Now consider the remaining alternative, that the p-optimal cyclic partition of $S$ includes a trivial cycle. We claim that in this case $\mathcal{D}_{\Sigma,\tina}^{[n,\nu]}=1.$ This is shown as follows. Suppose $\pi=\{i\}$ is a trivial cycle included in the p-optimal cyclic partition of $S$. Since the trivial cycle bound is active we must have $d_i=\alpha_{ii}=1$. Now, let User $j$ be  any other user in $S$. We immediately have the bound $d_i+d_j\leq \delta_{ij}+\delta_{ji}\leq 1$ (because in $\mathcal{N}^{[n,\nu]}$, all $\delta_{ij}\leq \nu/2$ and $\nu\leq 1$). Since $d_i=1$,  we must have $d_i+d_j=1$ and therefore, $d_j=0$. This is true for every user in $S$ besides user $i$. Therefore, $\mathcal{D}_{\Sigma,\tina}^{[n,\nu]}=1$ in this case.
\item The final step is to prove that $\mathcal{D}^{[n,\nu]}_{\Sigma,\tina}\leq 2$. Based on previous steps, this is proved as follows.
Isolating the left and right subnetworks of $\mathcal{N}^{[n,\nu]}$ from each other's interference does not hurt either of them, therefore,
\begin{align}
\mathcal{D}_{\Sigma,\tina}^{[n,\nu]}&\leq \mathcal{D}_{\Sigma,\tina}^{[n-1,\nu/2]}+\mathcal{D}_{\Sigma,\tina}^{[n-1,\nu/2]}\\
&=2 \mathcal{D}_{\Sigma,\tina}^{[n-1,\nu/2]}\\
&\leq 2\max\left(1, \frac{1}{2}\mathcal{D}_{\Sigma,\tina}^{[n-1,\nu]}\right)\\
&= \max(2, \mathcal{D}_{\Sigma,\tina}^{[n-1,\nu]})\\
&\leq\max(2,  \max(2, \mathcal{D}_{\Sigma,\tina}^{[n-2,\nu]}))\\
&=\max(2, \mathcal{D}_{\Sigma,\tina}^{[n-2,\nu]})\\
&\qquad \vdots \notag\\
&\leq\max(2,  \mathcal{D}_{\Sigma,\tina}^{[1,\nu]})\\
&= 2.
\end{align}
Thus, TIN cannot achieve more than $2$ sum-GDoF for our network. 
\end{enumerate}

Henceforth we will set $\nu=1$ and  prove that by allowing transmitter cooperation in this network, a sum-GDoF value of  $1+\frac{1}{2}\log_2(K)$ is achievable (and optimal). Recall that in a GDoF model, if Transmitter $j$ sends a message $W$ with power level $-\gamma_j$ to Receiver $i$ over a channel with strength $\alpha_{ij}$, then the received signal strength level is $\alpha_{ij}-\gamma_j$. The power levels are additive because these are exponents of $P$, or equivalently because they are being measured in dB scale. If the effective noise  floor, i.e., the maximum power level of noise and interference from other messages heard by Receiver $i$ is $\mu_i$, and $W$ carries $d_W$ GDoF, then $W$ can be decoded successfully while treating all other signals as noise if $d_W\leq \alpha_{ij}-\gamma_j-\mu_i$. Once a message is decoded it can be subtracted from the received signal before decoding other messages. This is the basic principle of successive decoding, and we will use it for the achievability proof. 

\begin{figure}[t]
	\centering
	\includegraphics[width=6.3in]{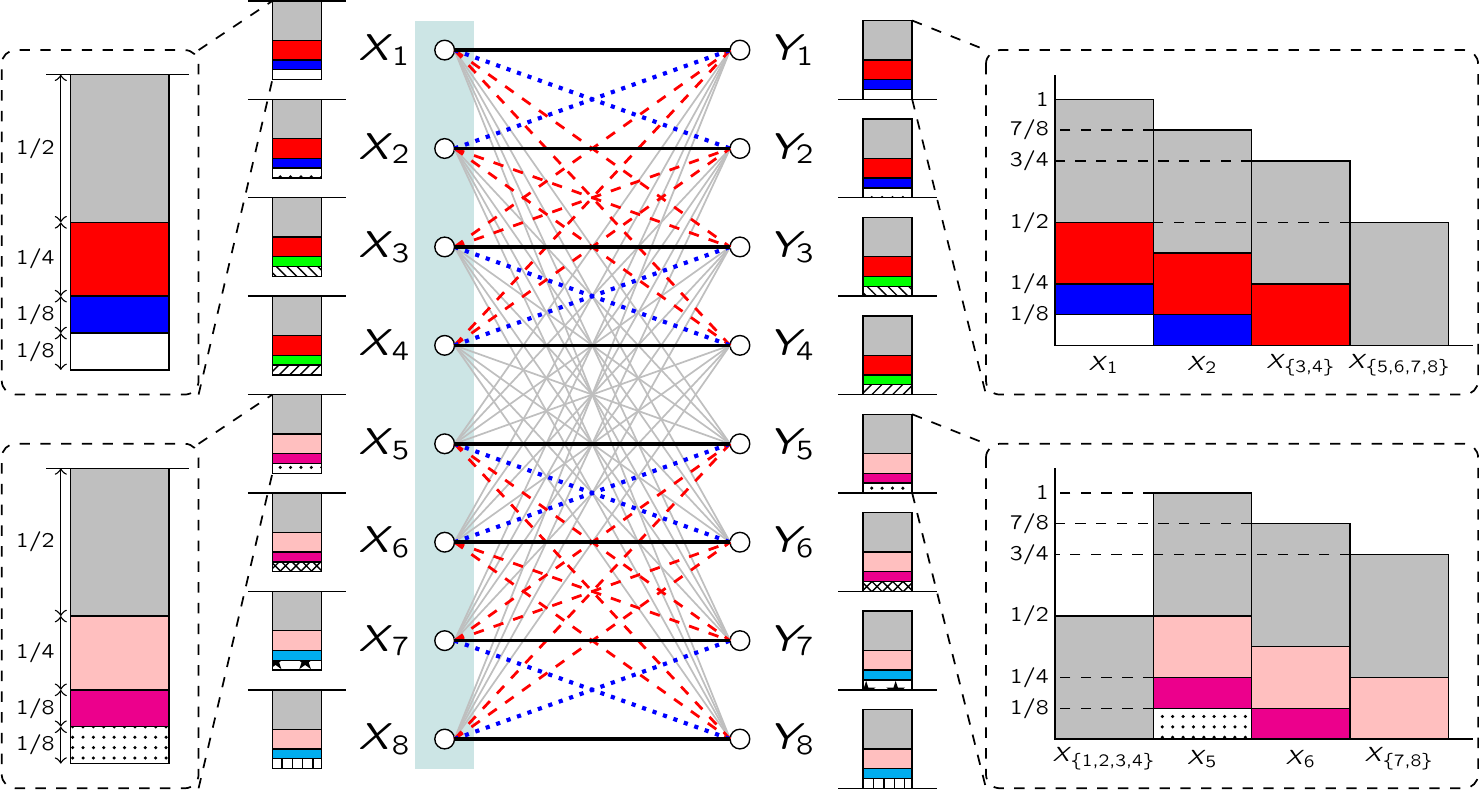}
	\caption{\small \it The SLS scheme for $\mathcal{N}^{[3,1]}$  that achieves $1+\frac{1}{2}\log_2(K)=\frac{5}{2}$ sum-GDoF by transmitter cooperation. }
	\label{fig:net}
\end{figure}

Before a detailed presentation of the achievable scheme for $\mathcal{N}^{[n,1]}$ networks, let us start with a sketch of  the achievable scheme for the example network $\mathcal{N}^{[3,1]}$, as shown in Fig. \ref{fig:net}. We saw the binary tree representation of this network earlier in Fig. \ref{fig:tree}. Recall that for this example, all direct links are of strength $\alpha_{ii}=1$. For the cross links, in Fig. \ref{fig:net} the dotted blue lines are links of strength $\alpha_{ij}=7/8$, the dashed red lines are of strength $\alpha_{ij}=3/4$, and the gray lines are links of strength $\alpha_{ij}=1/2$. The same gray common message at the top level is sent from all antennas to all users and carries $1/2$ sum GDoF. The dashed red links are in two separate clusters of $4$ users each, representing  $2$ subnetworks, each of the type $\mathcal{N}^{[2,1/2]}$ containing $4$ users. A red common message is sent for the first cluster and a pink common message is used for the second cluster, each carrying $1/4$ GDoF. Similarly, the dotted blue links are in $4$ separate clusters of $2$ users each, representing  $4$ subnetworks, each of the type $\mathcal{N}^{[1,1/4]}$ containing $2$ users. The corresponding blue, green, magenta and cyan power levels represent separate common messages for each of the $4$ subnetworks, carrying $1/8$ GDoF each. Finally, at the bottom level there is an independent message carrying $1/8$ GDoF for each user. The total sum-GDoF value thus achieved is $\left(\frac{1}{2}\right)+2\left(\frac{1}{4}\right)+4\left(\frac{1}{8}\right)+8\left(\frac{1}{8}\right)=5/2$. For the decoding, consider User $5$ as an example. The gray message which carries $1/2$ GDoF, is seen with power level $1$ and noise floor due to interference from other messages is at power level $1/2$ so it is successfully decoded and subtracted. Then  the pink message, which carries $1/4$ GDoF, is seen with power level $1/2$ and effective noise floor $1/4$, so it is also decoded and subtracted. Next,  the magenta message which carries $1/8$ GDoF is seen with power level $1/4$ and noise floor $1/8$, so it is also decoded and subtracted successfully. Finally, only the dotted white message, which carries $1/8$ GDoF is seen with power levels $1/8$ and noise floor $0$, so it is decoded as well. 

Now, let us explain the scheme for arbitrary $\mathcal{N}^{[n,1]}$. As in the example, the achievable scheme is also hierarchical where we will start with a common message for all users in $\mathcal{N}^{[n,\nu]}$ and then progressively include additional messages for its subnetworks while maintaining the successive decodability of all messages. For ease of reference, let us call the common message for the users in a $\mathcal{N}^{[n,\nu]}$ network a level $n$ message.

The same level-$n$ message, is sent from every transmitter with  strength $\gamma=0$, so that it is received at every receiver with strength $\gamma+\alpha_{ii}=1$.  It carries $0.5$ GDoF. The power levels of all other messages are set to $-1/2$ or less so that all other messages are received with strength no more than $-1/2+1=1/2$. Since the noise floor from other messages is at  $1/2$, the common message is received at strength level $1$, and it carries only $1/2$ GDoF, it is decodable at every receiver, After decoding it, every receiver subtracts out the codeword due to  the level $n$ message. 

There are two different level $n-1$ sub-networks. Within each of these two networks a different level $n-1$ message is sent with power level $-1/2$, so it is received at power level $1/2$ at each receiver within the sub-network. Signals from one sub-network are not heard by the other subnetwork because the channel strength between the users in different sub-networks is $1/2$ and the transmit power of the level $n-1$ message is $-1/2$. All lower level messages are sent with power levels less than $-3/4$, so the noise floor due to lower level messages at each receiver is at power level $1/4$. Thus, the level $n-1$ message is able to achieve $1/2-1/4=1/4$ GDoF. Since there are $2$ such messages corresponding to the $2$ subnetworks, the total sum GDoF value contributed by level $n-1$ messages is $1/4+1/4=1/2$. After decoding each receiver subtracts out the codeword due to level $n-1$ message from its own subnetwork.

Next, there are $4$ level $n-2$ sub-networks. A different common message is sent within each subnetwork with power level $-1+(1/2)^2=-3/4$, so it is received at power level $1/4$, while all lower level messages are sent with power no more than $-1+(1/2)^3=-7/8$, so the noise floor due to lower level messages is $1-7/8=1/8$. The sub-networks do not interfere with each other because the cross-subnetwork channel strengths are $1-1/2^2=3/4$ so the received signals from other subnetworks are below the noise floor. Thus, each of the $4$ of the $(n-2)$-level messages is able to achieve $1/4-1/8=1/8$ GDoF for a total of $4\times 1/8=1/2$. The decoded messages are subtracted.

This pattern continues, so that for each $i\in[0:n]$, there are $2^i$ different level-$(n-i)$ subnetworks. Within each of these subnetworks, a different common message is sent with power level $-1+(1/2)^i$ so it is received at power level $(1/2)^i$ while all lower level messages are sent with power level no more than $-1+(1/2)^{i+1}$ so that the noise floor due to lower level messages is $(1/2)^{i+1}$ at each receiver. Thus each of the $2^i$ subnetworks achieves $1/2^i - 1/2^{i+1}=1/2^{i+1}$ GDoF for a total of $2^i/2^{i+1}=1/2$ sum GDoF.

Adding these values across all $n$ levels we achieve a total of $n/2$ sum-GDoF. In fact, it is possible to do a little bit better. At level $0$, there are $2^n$ subnetworks comprised of individual users, and since there are no more lower level messages, the noise floor is $0$, so it is possible to achieve $1/2^n-0=1/2^n$ GDoF per user for a total of $1$ GDoF instead of just $1/2$ GDoF for level $0$ messages. Thus, the total sum-GDoF value achieved is $1+n/2=1+\frac{1}{2}\log_2(K)$ sum-GDoF. 
Now note that for the $\mathcal{N}^{[n,1]}$ network, the sum-GDoF value in the BC  setting is $\mathcal{D}_{\Sigma, \bc} \geq 1+\frac{1}{2}\log_2(K)$, while the sum-GDoF value achieved by TIN is $\mathcal{D}_{\Sigma, \tina} \leq 2$. Therefore, we have 
	\begin{align}
		\frac{\mathcal{D}_{\Sigma, \bc}}{\mathcal{D}_{\Sigma, \tina}} \geq \frac{1 + \frac{1}{2} \log_2(K)}{2} = \Theta( \log_2 (K)),
	\end{align}
which concludes the proof of the lower bound.

As a final remark, the sum-GDoF $1+\frac{n}{2}=1+\frac{1}{2}\log_2(K)$ is optimal for the BC obtained by allowing transmitter cooperation in $\mathcal{N}^{[n,1]}$. Applying Lemma \ref{lemma:bctocyc} with cycle $\pi = (1 \rightarrow 2 \rightarrow \cdots \rightarrow K \cyc~)$, we have the sum-GDoF in the BC bounded above by
\begin{align}
	\mathcal{D}_{\Sigma, \bc}^{[n,1]}([K]) &\leq \Delta_\pi + \alpha_{1K}^{[n,1]} \\
	&= \sum_{k=1}^{K-1} \delta_{k, k+1}^{[n, 1] }+\delta_{K1}^{[n,1]}+ \alpha_{1K}^{[n,1]} \\
	&=\sum_{\ell=1}^{n}\frac{1}{2^{n-\ell+1}}2^{n-\ell}+\frac{1}{2}+\frac{1}{2}\\
	&= \frac{1}{2} \log_2 K + 1,
\end{align}
which matches the achieved sum-GDoF.

\section{Conclusion}\label{sec:conc}
The results presented here open the door to a number of open questions where extremal analysis could be useful to gain a deeper understanding of the benefits of transmitter cooperation. For example, is it possible to achieve more than logarithmic GDoF gain by transmitter cooperation over TIN in a  \emph{general} weak interference regime where the only constraint is that the direct channels are stronger than cross channels?
What is the maximum possible sum-GDoF gain of a $K$ user MISO BC over the corresponding $K$ user IC in the general weak interference regime? Or, even in the SLS-regime? In general, it seems extremal analysis may be useful to gauge the  potential benefits of a myriad of factors such as multiple antennas, power control, rate-splitting, space-time multiplexing and network coherence -- all intriguing issues for which the current understanding is  extremely limited. Indeed, the main message of this work is to underscore the importance of extremal analysis in order to  advance our understanding of fundamental limits of large wireless networks beyond symmetric settings, where the curse of dimensionality  stands in the way. In particular, extremal analysis used in conjunction with the GDoF metric under finite precision CSIT, as exemplified by this work, appears to be a promising research avenue to bridge the gap between theory and practice.

\appendix
\section*{Appendix}
\section{Optimality of Cyclic Partition for Polyhedral TIN in  SLS Regime}\label{app:cycpart}
\begin{theorem}\label{theorem:main}\addcontentsline{toc}{subsection}{Theorem \ref{theorem:main}}
If $[\alpha]_{K\times K}\in\mathcal{A}_\sls$, then for any subset of users, $S$, $S\subset[K]$, there exists a p-optimal cyclic partition of $S$. 
\end{theorem}
\subsection{Proof of Theorem \ref{theorem:main}}
Without loss of generality we will prove the lemma for $S=[K]$, since the same proof works for any $S\subset[K]$ as well.
Let us start with arbitrary $[\alpha]_{K\times K}$, i.e., not necessarily in the SLS regime. The sum-GDoF value in the polyhedral region, $\mathcal{D}_{\Sigma,\ptin}$ is the solution to the following linear program,
\begin{align}
(LP_1)~~~~~~~  \mathcal{D}_{\Sigma}=\max & \mbox{ }d_1+d_2+\cdots+d_K\\
\mbox{such that}& \sum_{k\in\{\pi\}}d_k\leq \sum_{k\in\{\pi\}}\alpha_{kk}-w(\pi), ~~\forall \pi\in\Pi,\\
&d_k\geq 0, ~~ \forall k\in[K], \label{eq:positive}
\end{align}
and can be equivalently expressed by the following dual linear program. 

\begin{align}
(LP_2)~~~~~~~ \mathcal{D}_{\Sigma}=\min~ & \sum_{\pi\in\Pi}\lambda_\pi\left( \sum_{k\in\{\pi\}}\alpha_{kk}-w(\pi)\right)\\
\mbox{such that}&\sum_{\pi\in\Pi}\lambda_\pi 1(k\in\{\pi\})\geq1, ~~\forall k\in[K], \label{eq:unequal}\\
&\lambda_\pi\geq 0, ~~ \forall \pi\in\Pi,
\end{align}
where $1(\cdot)$ is the indicator function that returns the values 1 or 0 when the argument to the function is true or false, respectively.

For all $\pi\in\Pi$, let us define $\lambda_\pi^*$ as the optimizing values of $\lambda_\pi$ for $LP_2$. Let the corresponding optimal values for $LP_1$ be $d_k^*$ for all $k\in[K]$. Because a solution must exist, by the strong-duality of linear programming, the optimal $D_\Sigma$ for  $LP_2$ is the same as the optimal $D_\Sigma$ for  $LP_1$.  Therefore, the following conditions are implied.
\begin{align}
D_\Sigma=d_1^*+d_2^*+\cdots+d_K^*&=\sum_{\pi\in\Pi}\lambda_\pi^*\left(\sum_{k\in\{\pi\}}\alpha_{kk}-w(\pi)\right),\\
 \sum_{k\in\{\pi\}}\alpha_{kk}-w(\pi)&\geq \sum_{k\in\{\pi\}}d^*_k, &\forall \pi\in[\Pi],\\
\lambda_\pi^*&\geq 0, &\forall \pi\in[\Pi],\\
d_k^*&\geq 0, &\forall k\in[K].
\end{align}

\begin{definition}[Set of Active Cycles, $\Pi^*$]
Based on the optimizing solution to $LP_2$, define
\begin{align}
\Pi^*&=\{\pi\in\Pi: \lambda^*_\pi>0\}.
\end{align}
\end{definition}
This is called the set of active cycles, because the corresponding cycle bounds are active (i.e., tight) in the solution to $LP_2$ (see Lemma \ref{lemma:inactive}).
\begin{definition}[Set of Inactive Users, $\mathcal{K}_o$] Define $\mathcal{K}_o\subset [K]$ as the set of all users $k$ for which the inequality in \eqref{eq:unequal} is strict. Thus, 
\begin{align}
\mathcal{K}_o&=\{k\in[K]:\sum_{\pi\in\Pi}\lambda^*_\pi 1(k\in\{\pi\})>1 \}.
%k\in\mathcal{K}_o&&\Rightarrow&&\sum_{\pi\in\Pi}\lambda^*_\pi 1(V_k\in\pi)>1.
\end{align}
\end{definition}
This is called the set of inactive users because for each of these users, we must have $d_k^*=0$ (see Lemma \ref{lemma:inactive}).
%\begin{mdframed}[backgroundcolor=black!4]
\begin{lemma}\label{lemma:inactive}
\begin{align}
\forall k\in\mathcal{K}_o &&\mbox{ we must have } && d_k^*=0,\\
\mbox{\normalfont and } \forall\pi\in\Pi^*&&\mbox{ we must have }&& \sum_{k\in\{\pi\}}d_k^*&=\sum_{k\in\{\pi\}}\alpha_{kk}-w(\pi).
\end{align}
\end{lemma}
%\end{mdframed}
 Note that the conditions are simply complementary slackness conditions, therefore Lemma \ref{lemma:inactive} holds for arbitrary channel parameters, i.e., even if $[\alpha]_{K\times K}\notin \mathcal{A}_{\mbox{\tiny SLS}}.$ For the sake of completeness, a proof of Lemma \ref{lemma:inactive} appears in Appendix \ref{app:inactive}.

Henceforth, let us restrict our attention to the SLS regime. In fact, let us define a \emph{strict} SLS regime as 
\begin{align}
\bar{\mathcal{A}}_{\mbox{\tiny SLS}}=\{[\alpha]_{K\times K}\in\mathbb{R}_+^{K\times K}: \alpha_{ii}> \max(\alpha_{ij},\alpha_{ki}, \alpha_{ik}+\alpha_{ji} -\alpha_{jk}), ~\forall i,j,k\in[K], i\notin\{j,k\}\}.
\end{align}
Note that the only difference between $\mathcal{A}_\sls$ and $\bar{\mathcal{A}}_\sls$ is that the defining inequalities in the latter are all strict inequalities. Note that all $\alpha_{ii}$ and $\delta_{ij}$ are strictly positive in the strict SLS regime. Following the same reasoning as the proof of Lemma \ref{lemma:delta}, in the strict SLS regime, for distinct $i,j,k\in[K]$, we must have
\begin{align}
[\alpha]_{K\times K}\in \bar{\mathcal{A}}_{\mbox{\tiny SLS}}&&\Rightarrow && \delta_{ki}+\delta_{ij}&>\delta_{kj}. \label{eq:slsdeltabar}
\end{align}
Note that the inequality is strict here as well. This is important for the proof.

We will first prove Theorem \ref{theorem:main} for the strict SLS regime and  later use a continuity argument (identical to the continuity argument in the last paragraph of the proof of Theorem 3 in \cite{Sun_Jafar_ParallelTIN}) to show that the result holds even when the inequalities are relaxed to include equalities. The shell of the proof is identical to the proof of Theorem 3 in \cite{Sun_Jafar_ParallelTIN}. The main  step that connects the two proofs is Lemma \ref{lemma:key} in this paper.

Now define the following linear program.
\begin{align}
(LP_3)~~~~~~~ \mathcal{D}_\Sigma=\min~ & \sum_{\pi\in\Pi}\lambda_\pi\left( \sum_{k\in\{\pi\}}\alpha_{kk}-w(\pi)\right)\\
\mbox{such that}&\sum_{\pi\in\Pi}\lambda_\pi 1(k\in\{\pi\})=1, ~~\forall k\in[K], \label{eq:equal}\\
&\lambda_\pi\geq 0, ~~ \forall \pi\in\Pi.
\end{align}
Note that the only difference between $LP_2$ and $LP_3$ is that the inequality in \eqref{eq:unequal} has been replaced with the equality in \eqref{eq:equal}. The following lemma is the most critical part of the proof, as it shows that this change does not matter in the strict SLS regime, thereby reducing the problem to another problem that is already solved in \cite{Sun_Jafar_ParallelTIN}.
\begin{lemma} \label{lemma:key}
\begin{align}
[\alpha]_{K\times K}\in \bar{\mathcal{A}}_{\mbox{\tiny SLS}}&&\Rightarrow&& LP_2\equiv LP_3.
\end{align}
\end{lemma}

The proof of Lemma \ref{lemma:key} appears in Appendix \ref{app:key}.

\noindent
Following Lemma \ref{lemma:key}, $LP_3$ is identical to the $LP_3$ in  \cite{Sun_Jafar_ParallelTIN} and the rest of the proof is identical to the proof of Theorem 3 in \cite{Sun_Jafar_ParallelTIN}. Thus, the proof of Theorem \ref{theorem:main} is complete. $\hfill\square$

{\it Remark:  Note that Lemma \ref{lemma:key} does not follow from \cite{Sun_Jafar_ParallelTIN}. Only \underline{after} Lemma \ref{lemma:key} do the two proofs become identical. In \cite{Sun_Jafar_ParallelTIN}, the equivalence of $LP_2$ and $LP_3$ is proved for a strict TIN regime. However, that proof does not hold in the  strict SLS regime, and this distinction is quite important. In both cases (strict TIN regime and the strict SLS regime), we need to prove that all the constraints in \eqref{eq:unequal} are tight. In the strict TIN regime, \cite{Sun_Jafar_ParallelTIN} accomplishes this by first proving that all $d_i^*$ that optimize the sum-GDoF must be strictly positive, so that it follows from complementary slackness  that the constraints in  \eqref{eq:unequal} must be tight. However, in the strict SLS regime, unfortunately it is not true that all $d_i^*$ \emph{must} be strictly positive. A simple counterexample is the $2$ user IC with $\alpha_{11}=\alpha_{22}=1, \alpha_{12}=\alpha_{21}=1/2$ which is in the strict SLS regime but not the strict TIN regime, and has $\mathcal{D}_{\Sigma,\tina}=1$ which can be achieved with $(d_1^*,d_2^*)=(1,0)$. Therefore, Lemma \ref{lemma:key} in the strict SLS regime needs a different argument that proves directly that all conditions in \eqref{eq:unequal} are tight without relying on strict positivity of all the $d_i^*$ that optimize the sum-GDoF. Such an argument is presented in Appendix \ref{app:key}.}

\subsection{Proof of Lemma \ref{lemma:inactive}}\label{app:inactive}
\begin{align}
0&=\sum_{\pi\in\Pi}\lambda_\pi^*\left(\sum_{k\in\{\pi\}}\alpha_{kk}-w(\pi)\right)-D_\Sigma\label{eq:step1}\\
&\geq \sum_{\pi\in\Pi}\lambda_\pi^*\left(\sum_{ k \in\{\pi\}}d_k^*\right)-D_\Sigma\\
&=\sum_{k\in[K]}\left(d_k^*\left(\sum_{\pi\in[\Pi]}\lambda_\pi^*1(k \in\{\pi\})\right)\right)-D_\Sigma\\
&=\sum_{k\in[K]\backslash\mathcal{K}_o}d_k^*+\sum_{k\in\mathcal{K}_o}c_kd_k^*-D_\Sigma\\
&=\sum_{k\in[K]\backslash\mathcal{K}_o}d_k^*+\sum_{k\in\mathcal{K}_o}c_kd_k^*-\sum_{k\in[K]}d_k^*\\
&=\sum_{k\in\mathcal{K}_o}(c_k-1)d_k^*\\
&\geq 0, \label{eq:step7}
\end{align}
 because $c_k\triangleq\sum_{\pi\in\Pi}\lambda_\pi^*1( k \in\pi)>1$ for all $k\in\mathcal{K}_o$, and $d_k^*\geq 0$ for all $k\in[K]$. Since we started and ended with $0$, all steps from \eqref{eq:step1} to \eqref{eq:step7} must be equalities. Thus, the proof of Lemma \ref{lemma:inactive} is complete.$\hfill\square$

\subsection{Proof of Lemma \ref{lemma:key}}\label{app:key}
We need to prove that the set $\mathcal{K}_o$ is empty.  Suppose, on the contrary, that there exists $k_o\in\mathcal{K}_o$. According to Lemma \ref{lemma:inactive} the user $k_o$ must be inactive, i.e., $d^*_{k_o}=0$. Let $\pi_o=(i_1\rightarrow i_2\rightarrow\cdots\rightarrow i_M\cyc~)$ be an active cycle that includes User ${k_o}$. Without loss of generality, suppose $k_o=i_M$. We will consider $3$ cases.
\begin{enumerate}
\item{\bf Case 1: $(M>2)$}\\
Suppose the length of the cycle is greater than $2$. Since $\pi_o$ is an active cycle, according to Lemma \ref{lemma:inactive},
\begin{align}
\sum_{k \in\{\pi_o\}}d_k^*&=\sum_{ k \in\{\pi_o\}}\alpha_{kk}-w(\pi_o)\\
\implies d_{i_1}^*+d_{i_2}^*+\cdots+d_{i_M}^*&=\delta_{i_1i_2}+\delta_{i_2i_3}+\cdots+\delta_{i_{M-1}i_M}+\delta_{i_Mi_1}.
\end{align}
But since $k_o\in\mathcal{K}_o$, according to Lemma \ref{lemma:inactive} we must have $d_{k_o}^*=d_{i_M}^*=0$. Therefore,
\begin{align}
d_{i_1}^*+d_{i_2}^*+\cdots+d_{i_{M-1}}^*&=\delta_{i_1i_2}+\delta_{i_2i_3}+\cdots+\delta_{i_{M-2}i_{M-1}}+\delta_{i_{M-1}i_M}+\delta_{i_Mi_1}.\label{eq:cyctight}
\end{align}
But now consider the cycle $\pi'=(i_1\rightarrow i_2\cdots\rightarrow i_{M-1}\cyc~)$. This may or may not be an active cycle. Regardless, the following bound must hold.
\begin{align}
d_{i_1}^*+d_{i_2}^*+\cdots+d_{i_{M-1}}^*&\leq \delta_{i_1i_2}+\delta_{i_2i_3}+\cdots+\delta_{i_{M-2}i_{M-1}}+\delta_{i_{M-1}i_1}. \label{eq:cycloose}
\end{align}
Subtracting \eqref{eq:cyctight} from \eqref{eq:cycloose} we have
\begin{align}
0&\leq \delta_{i_{M-1}i_1}-\delta_{i_{M-1}i_M}-\delta_{i_Mi_1}\\
\implies \delta_{i_{M-1}i_M}+\delta_{i_Mi_1}&\leq \delta_{i_{M-1}i_1}.
\end{align}
But this is a contradiction because under strict SLS condition, according to  \eqref{eq:slsdeltabar},
\begin{align}
\delta_{i_{M-1}i_M}+\delta_{i_Mi_1}&>\delta_{i_{M-1}i_1}.
\end{align}

\item{\bf Case 2: $(M=1)$}\\
The length of the cycle, $M$, cannot be $1$ because then Lemma \ref{lemma:inactive} would imply that the single user bound is active, i.e., $d^*_{k_o}=\alpha_{k_ok_o}$, but $\alpha_{k_ok_o}>0$ in the strict SLS regime, so  user $k_o$ must be active, i.e., we would have a contradiction.  This leaves us with the only possibility, $M=2$.

\item{\bf Case 3: $(M=2)$}\\
Now suppose the length of the cycle $\pi_o$ is $M=2$. Then we have
\begin{align}
d_{i_1}^*+d_{i_M}^*&=\delta_{i_1i_M}+\delta_{i_Mi_1},
\end{align}
and since $k_o=i_M\in\mathcal{K}_o$ according to Lemma \ref{lemma:inactive} we have $d_{i_M}^*=0$. Therefore,
\begin{align}
d_{i_1}^*&=\delta_{i_1i_M}+\delta_{i_Mi_1}. \label{eq:a}
\end{align}

Consider the following two subcases.
\begin{enumerate}
\item {\bf Subcase 1: $\pi_o$ is the only active bound that includes user $k_o$} \\
Then $\lambda_{\pi_o}>1$. But this would mean that user $i_1$ also belongs to $\mathcal{K}_o$, because the sum of weights of active cycles that include user $i_1$ must be greater than $1$ as well. However, if both user $i_1$ and user $i_M$ are in $\mathcal{K}_o$, then they must both be inactive. This is a contradiction, because $d_{i_1}^*+d_{i_M}^*=\delta_{i_1i_M}+\delta_{i_Mi_1}>0$.
\item  {\bf Subcase 2: There is another active bound, $\pi_1\neq \pi_o$  that includes user $k_o$}\\
Now, $\pi_1$ must also have length $M=2$ because, as we have already established, any other possibility leads to a contradiction. Since $\pi_1$ is different from $\pi_o$ it must involve a user other than $i_1$ in addition to user $i_M$. Let's call this user $i_2$. Then, proceeding similarly as in the case of $\pi_o$ we find that we must have
\begin{align}
d_{i_2}^*&=\delta_{i_2i_M}+\delta_{i_Mi_2}. \label{eq:b}
\end{align}
But we also know that the following bound must hold
\begin{align}
d_{i_1}^*+d_{i_2}^*&\leq \delta_{i_1i_2}+\delta_{i_2i_1}.  \label{eq:c}
\end{align}
Subtracting \eqref{eq:a} and \eqref{eq:b} from \eqref{eq:c} we have,
\begin{align}
0&\leq\delta_{i_1i_2}+\delta_{i_2i_1}-\delta_{i_1i_M}-\delta_{i_Mi_1}-\delta_{i_2i_M}-\delta_{i_Mi_2}\label{eq:usedelta}\\
&<(\delta_{i_1i_M}+\delta_{i_Mi_2})+(\delta_{i_2i_M}+\delta_{i_Mi_1})-\delta_{i_1i_M}-\delta_{i_Mi_1}-\delta_{i_2i_M}-\delta_{i_Mi_2}\\
&=0,
\end{align}
which is a contradiction. Note that we used  \eqref{eq:slsdeltabar} in \eqref{eq:usedelta}.
\end{enumerate}
\end{enumerate}

Thus, we have a contradiction in every case, so there cannot be any such $k_o\in\mathcal{K}_o$, which implies that $\mathcal{K}_o$ is empty, and the proof is complete.$\hfill\square$

\section{Other Useful Lemmas}\label{app:useful}
\subsection{A condition on $\delta_{ij}$ in the SLS Regime}
\begin{lemma}\label{lemma:delta} 
For all  $i,j,k\in[K]$,
\begin{align}
[\alpha]_{K\times K}\in {\mathcal{A}}_{\mbox{\tiny SLS}}&&\Rightarrow && \delta_{ki}+\delta_{ij}&\geq \delta_{kj}. \label{eq:slsdelta}
\end{align}
\end{lemma}
\subsection*{Proof of Lemma \ref{lemma:delta}}
{\it Proof:}
$\delta_{ki}+\delta_{ij}-\delta_{kj}=\alpha_{kk}-\alpha_{ik}+\alpha_{ii}-\alpha_{ji}-\alpha_{kk}+\alpha_{jk}=\alpha_{ii}+\alpha_{jk}-\alpha_{ik}-\alpha_{ji}$ which, by definition, is non-negative in $\mathcal{A}_\sls$. $\hfill\square$
 
\subsection{Trivial cycles in the SLS Regime}
\begin{lemma}\label{lemma:trivial}
	If $[\alpha]_{K \times K} \in \mathcal{A}_\sls$,  then for every $\mathcal{S} \subset [K]$ there exists a p-optimal cyclic partition containing at most one trivial cycle.
\end{lemma}
\subsection*{Proof of Lemma \ref{lemma:trivial}}
Let $\{ \pi_i \}_{i=1}^{N}$ be a p-optimal cyclic partition for $\mathcal{S}$. 
Suppose there is more than one trivial cycle in a p-optimal cyclic partition, we claim that they can be combined into one cycle, and the resulting partition is still p-optimal and free of trivial cycles.
Let $\pi_1 = (i_1 \cyc~), \pi_2 = (i_2\cyc~), \cdots, \pi_j = (i_j \cyc~)$, $ 2 \leq j \leq N$, be all the trivial cycles in $\{ \pi_i \}_{i=1}^{N}$. 
These trivial cycles can be combined into $\pi_{1,2,\cdots,j} = (\pi_1 \rightarrow \pi_2 \rightarrow \cdots \rightarrow \pi_j \cyc~)$.
Since $\pi_{1,2, \cdots, j}$ and all the other cycles are disjoint, $\{ \pi_{1,2, \cdots, j}, \pi_{j+1}, \cdots, \pi_N  \}$ is a cyclic partition. 
Moreover, 
\begin{align}
\Delta_{\pi_{1,2,\cdots,j}} 
= \sum_{m=1}^{j} \delta_{i_{m}, i_{m+1}} 
\leq \sum_{m=1}^{j} \alpha_{i_m i_m} = \sum_{m=1}^{j} \Delta_{\pi_m},
\end{align}
where $\delta_{i_{j}, i_{j+1}} = \delta_{i_{j}, i_{1}}$.
As a result, $\{\pi_{1,2,\cdots,j}, \pi_{j+1}, \cdots, \pi_N \}$ is also p-optimal, and contains no trivial cycles. $\hfill\square$
 
\subsection{Combining Disjoint Cycles in the SLS Regime}
\begin{lemma}\label{lemma:joincycle}
If $[\alpha]_{K\times K}\in\mathcal{A}_\sls$,  $\pi_1,\pi_2,\cdots,\pi_n$ are $n>1$ disjoint cycles, and 
\begin{align}
\pi_{1,2,\cdots,n}&=(\pi_1\rightarrow\pi_2\rightarrow\cdots\rightarrow\pi_n\cyc~)
\end{align} 
is their combination, then
\begin{align}
\Delta_{\pi_{1,2,\cdots,n}}&\leq \Delta_{\pi_1}+\Delta_{\pi_2}+\cdots+\Delta_{\pi_n}+\Delta_\pi,
\end{align}
where $\pi=(\pi_1(1)\rightarrow\pi_2(1)\rightarrow\cdots\rightarrow\pi_n(1)\cyc~)$.
\end{lemma}
%\section{Proofs of Lemmas}
\subsection*{Proof of Lemma \ref{lemma:joincycle}}
Let us represent the cycles explicitly as
\begin{align}
\pi_1&=(i_{1,1}\rightarrow\cdots\rightarrow i_{1,m_1}\cyc~),\\
\pi_2&=(i_{2,1}\rightarrow\cdots\rightarrow i_{2,m_2}\cyc~),\\
&\vdots\notag\\
\pi_n&=(i_{n,1}\rightarrow\cdots\rightarrow i_{n,m_n}\cyc~),\\
\pi_{1,2,\cdots,n}&=(i_{1,1}\rightarrow\cdots\rightarrow i_{1,m_1}\rightarrow i_{2,1}\rightarrow\cdots\rightarrow i_{2,m_2}\rightarrow\cdots\rightarrow i_{n,m_n}\cyc~).
\end{align}
Then we have
\begin{align}
\Delta_{\pi_{1,2,\cdots,n}}&\leq (\Delta_{\pi_1}-\delta_{i_{1,m_1}i_{1,1}})+(\Delta_{\pi_2}-\delta_{i_{2,m_2}i_{2,1}})+\cdots+(\Delta_{\pi_n}-\delta_{i_{n,m_n}i_{n,1}})\notag\\
&\hspace{1cm}+\delta_{i_{1,m_1}i_{2,1}}+\delta_{i_{2,m_2}i_{3,1}}+\cdots+\delta_{i_{{n-1},m_{n-1}}i_{n,1}}+\delta_{i_{n,m_n}i_{1,1}}\\
&\leq \Delta_{\pi_1}+\delta_{i_{1,1}i_{2,1}}+\Delta_{\pi_2}+\delta_{i_{2,1}i_{3,1}}+\cdots+\Delta_{\pi_n}+\delta_{i_{n,1}i_{1,1}}\label{eq:usesls}\\
&= \Delta_{\pi_1}+\Delta_{\pi_2}+\cdots+\Delta_{\pi_n}+\Delta_\pi.
\end{align}
Note that in \eqref{eq:usesls} we used the fact that since $[\alpha]_{K\times K}\in\mathcal{A}_\sls$, we must have $\delta_{ij}+\delta_{jk}\geq \delta_{ik}$. $\hfill\square$

\subsection{Connecting BC Bounds to Cycle Bounds in the SLS Regime}\label{sec:bctocyc}
\begin{lemma}\label{lemma:bctocyc}
In the SLS regime, for any cycle $\pi\in\Pi$, 
\begin{align}
\pi&=(i_1\rightarrow i_2\rightarrow\cdots\rightarrow i_M\cyc~),
\end{align}
we have the following bound on the sum-GDoF of the BC restricted to the users involved in the cycle $\pi$,
\begin{align}
\mathcal{D}_{\Sigma,\bc}(\{\pi\})&\leq \Delta_\pi+\alpha_{i_{m+1}i_m},
%\begin{cases}
%\Delta_\pi & \text{, if } M = 1 \\
% & \text{, if } M \geq 2 
%\end{cases}
\end{align}
for any $m\in[1:M]$, with $i_{M+1}=i_1$. Furthermore,
\begin{align}
\mathcal{D}_{\Sigma,\bc}(\{\pi\})&\leq \Delta_\pi+\mathcal{D}_{\Sigma,\tina}.
\end{align}
\end{lemma}
\subsection*{Proof of Lemma \ref{lemma:bctocyc}}
Lemma \ref{lemma:bctocyc} follows directly as a special case of the results presented in \cite{Arash_Jafar_SLS}. For the sake of completeness we present a self-contained proof here. The proof is trivial for cycles of length $M=1$, because the single-user bound implies $\mathcal{D}_{\Sigma, \bc} (\{ \pi \}) \leq \alpha_{i_1i_1} \leq  \Delta_\pi + \alpha_{i_1i_1}$. To prove Lemma \ref{lemma:bctocyc} for $M\geq 2$, let us give to each receiver $i_m,m\in[M]$, the messages $W_{[m+1:M]}\triangleq(W_{i_{m+1}}, W_{i_{m+2}},\cdots, W_{i_M})$ as side information. This can only help, so the converse for the genie-aided channel is still a converse for the original channel. Note that no messages are given as side information  to receiver $M$. Now, 
applying Fano's inequality within the deterministic model (Section \ref{sec:det}) of the $K$ user MISO broadcast channel, and omitting  $o(\log (P))$ terms  we have,
\begin{align}
TR_{i_1}&\leq I( W_{i_1}; (\bar{Y}_{i_1}(t))^{[1:T]} | \mathcal{G} , W_{[2:M]}) \\
&\leq H( (\bar{Y}_{i_1}(t))^{[1:T]} | \mathcal{G}, W_{[2:M]}) \\
TR_{i_m}&\leq I( W_{i_m}; (\bar{Y}_{i_m}(t))^{[1:T]} | \mathcal{G} , W_{[m+1:M]}) \\
&= H((\bar{Y}_{i_m}(t))^{[1:T]} | \mathcal{G} , W_{[m+1:M]}) -H((\bar{Y}_{i_m}(t))^{[1:T]} | \mathcal{G} , W_{[m:M]}), &\forall m\in[2:M].
%T R_{\mu(i)} &\leq I( W_{\mu(i)}; (\bar{Y}_{\mu(i)}(t))^{[1:T]} | \mathcal{G} ) \\
%&\leq I( W_{\mu(i)}; (\bar{Y}_{\mu(i)}(t))^{[1:T]} | \mathcal{G}, W_{\mu(1)}, W_{\mu(i+1)}, \cdots, W_{\mu(M)} ) \label{eq:indepZ1}\\
%&= H((\bar{Y}_{\mu(i)}(t))^{[1:T]} |\mathcal{G}, W_{\mu(1)}, W_{\mu(i+1)}, \cdots, W_{\mu(M)} ) \notag \\
%& \qquad - H( (\bar{Y}_{\mu(i)}(t))^{[1:T]} |\mathcal{G}, W_{\mu(1)}, W_{\mu(i)}, W_{\mu(i+1)}, \cdots, W_{\mu(M)}  ), &&\text{for $i \in [3:M]$}\\
%T R_{\mu(2)} &\leq I( W_{\mu(2)}; (\bar{Y}_{\mu(2)}(t))^{[1:T]} | \mathcal{G} ) \\
%&\leq I( W_{\mu(2)}; (\bar{Y}_{\mu(2)}(t))^{[1:T]} | \mathcal{G}, W_{\mu(1)}, W_{\mu(3)},\cdots, W_{\mu(M)} )\label{eq:indepZ2}\\
%&= H((\bar{Y}_{\mu(2)}(t))^{[1:T]} |\mathcal{G},W_{\mu(1)}, W_{\mu(3)}, \cdots, W_{\mu(M)} ) \label{eq:fullW}
\end{align}
%In (\ref{eq:indepZ1}) and (\ref{eq:indepZ2}), we use the fact that, for the random variables $X,Y,Z$, $I(X;Y) \leq I(X;Y|Z)$ if $Z$ is independent of $X$ or $Y$.
%The indices for the cycle of length $M$ are modulo $M$ in (\ref{eq:indepZ1}), i.e., $\mu(M+1) = \mu(1)$.
%(\ref{eq:fullW}) holds because, given all messages $\{ W_\mu(i)  \}_{i=1}^{M}$ and $\mathcal{G}$, one can recover $\bar{X}_{\mu(i)}(t)$ for all $i \in [1:M]$ and $t \in [1:T]$, and thus $\bar{Y}_{\mu(M)}(t)$ for all $t \in [1:T]$. 

Adding these inequalities we get,
\begin{align}
&T \sum_{m=1}^{M} R_{i_m} \notag\\
\leq&  \sum_{m=1}^{M-1} \left[ H\left((\bar{Y}_{i_m}(t))^{[1:T]} |\mathcal{G}, W_{[m+1:M]} \right)-H\left((\bar{Y}_{i_{m+1}}(t))^{[1:T]} |\mathcal{G}, W_{[m+1:M]}\right)\right]+H( (\bar{Y}_{i_M}(t))^{[1:T]} | \mathcal{G})  \\
\leq & \sum_{m=1}^{M-1} \Big [  \mathbb{H}_g \left( [\alpha_{i_m1},   \alpha_{i_m2}, \cdots,  \alpha_{i_mK}] \mid W_{[m+1:M]}  \right)   - \mathbb{H}_g\left ( [\alpha_{i_{m+1}1},   \alpha_{i_{m+1}2}, \cdots,  \alpha_{i_{m+1}K} ]\mid W_{[m+1:M]} \right ) \Big ] \notag\\
&\hspace{1cm}+\alpha_{i_Mi_M}T\log(P) \\
\leq&\sum_{m=1}^{M-1}\max_{\ell\in [K]}\left(\alpha_{i_m\ell}-\alpha_{i_{m+1}\ell}\right)^+T\log(P)+\alpha_{i_Mi_M}T\log(P) \label{eq:maxell}\\
\leq&\sum_{m=1}^{M-1}\left(\alpha_{i_mi_m}-\alpha_{i_{m+1}i_m}\right)T\log(P)+\alpha_{i_Mi_M}T\log(P) \label{eq:bcSLS}\\
\leq&\sum_{m=1}^{M}\left(\alpha_{i_mi_m}-\alpha_{i_{m+1}i_m}\right)T\log(P)+\alpha_{i_1i_M}T\log(P) \\
=&\left(\sum_{m=1}^{M}\delta_{i_mi_{m+1}}\right)T\log(P)+\alpha_{i_1i_M}T\log(P) \\
=&(\Delta_\pi +\alpha_{i_1i_M})T\log(P).  \label{eq:last}
\end{align}
Note that  Lemma \ref{lemma:AIS} was used in  \eqref{eq:maxell}, and the definition of the SLS regime was used in \eqref{eq:bcSLS} to conclude that
\begin{align}
\alpha_{i_mi_m} - \alpha_{ i_{m+1}i_m} & \geq \alpha_{i_m,\ell} - \alpha_{ i_{m+1},\ell}.
\end{align}
From \eqref{eq:last} we have in the GDoF limit,
\begin{align}
\mathcal{D}_{\Sigma, \bc} (\{ \pi  \})  \leq  \Delta_\pi + \alpha_{i_{1}i_{M}}= \Delta_\pi + \alpha_{\pi(M+1)\pi(M)}.
\end{align}
Next, note that if we go through the same steps starting with a shifted representation of the cycle $\pi$, e.g., 
\begin{align}
\pi^j&=(\pi(1+j)\rightarrow\pi(2+j)\rightarrow\cdots\rightarrow\pi(M+j)\cyc),
\end{align}
then we obtain 
\begin{align}
\mathcal{D}_{\Sigma, \bc} (\{ \pi^j  \}) =\mathcal{D}_{\Sigma, \bc} (\{ \pi \})  \leq  \Delta_\pi + \alpha_{\pi(M+1+j)\pi(M+j)},
\end{align}
and in particular for $j=m+M$, we have the bound $\mathcal{D}_{\Sigma, \bc} (\{ \pi \})  \leq\Delta_\pi + \alpha_{\pi(m+1)\pi(m)}=\Delta_\pi + \alpha_{i_{m+1}i_{m}}$ for any $m\in[1:M]$, as desired. Finally, seeing that $\alpha_{ij} \leq \alpha_{ii} \leq \mathcal{D}_{\Sigma, \tina}(\{\pi \})$ for all $i, j \in \{\pi \}$ in the SLS regime, we have 
\begin{align}
\mathcal{D}_{\Sigma, \bc}(\{\pi \}) \leq \Delta_\pi + \mathcal{D}_{\Sigma, \tina}(\{\pi \}).
\end{align}
This completes the proof of Lemma \ref{lemma:bctocyc}.$\hfill\square$

\end{document}